%% file: main.tex
\documentstyle[prd,aps]{revtex}
\input makro1
\input makro2
\begin{document}
\draft
\title{Symmetries of asymptotically flat electrovacuum spacetimes
\protect\\
%admitting gravitational and electromagnetic
and  radiation}
\author{J.~Bi\v c\' ak\footnote{Email address: 
{\tt bicak@mbox.troja.mff.cuni.cz} }  
\ and A.~Pravdov\' a\footnote{Email address: 
{\tt pravdova@otokar.troja.mff.cuni.cz}}\\}
\address{Department of Theoretical Physics, Faculty of Mathematics
and Physics, \protect\\
Charles University, V Hole\v sovi\v ck\' ach
2,\protect\\
180 00 Prague 8, Czech Republic}
\date{\today}
\maketitle
\begin{abstract}
Symmetries compatible with asymptotic flatness and
admitting gravitational and electromagnetic radiation
are studied %in physical spacetime
by using
the~Bondi-Sachs-van der Burg formalism. It is  shown that
in axially symmetric electrovacuum spacetimes in which
at least locally a smooth null infinity in the~sense
of Penrose exists, the~only second allowable symmetry
is either the~translational symmetry or the~boost
symmetry. Translationally invariant spacetimes
with in general a straight "cosmic string"  along
the~axis of symmetry are non-radiative although
they can have a non-vanishing news function.
The~boost-rotation symmetric spacetimes are radiative.
They describe "uniformly accelerated charged particles"
or black holes which in general may also be rotating -
the~axial and an additional Killing vector are not assumed to
be hypersurface orthogonal.
The~general functional forms of both gravitational
and electromagnetic news functions, and of the~mass aspect
and total mass of asymptotically flat boost-rotation
symmetric spacetimes at null infinity are obtained. The~expressions
for the~mass are new even in  the~case of vacuum
boost-rotation symmetric spacetimes with hypersurface
orthogonal Killing vectors. In Appendices some errors
appearing in previous works are corrected.
\end{abstract}
%$\verb+\pacs{#1}+ command
%{\tt$\backslash$\string pacs\{\}} should always be input,
%even if empty.}

%\narrowtext

\input ajkap1
\input ajkap2
\input ajkap3
\input ajkap4
\\[2mm]
\begin{center}
{\bf{ACKNOWLEDGMENTS}}\\
\end{center}
We thank Vojt\v ech Pravda for discussions and help with
calculations. We are grateful to Piotr Chru\' sciel for reading
the manuscript and making helpful suggestions. We also 
acknowledge support from grants GACR-202/96/0206
of the Czech Republic and GAUK-230/96 of the~Charles University,
and the hospitality of the Albert Einstein Institute, Potsdam.
\appendix
\section{The~Einstein-Maxwell equations\protect\\
 in the~Bondi-Sachs coordinates}
\input ajap1
\section{The~Killing equation}
\input ajap2
\section{Corrections of Killing equations given in Ref. [2]}
\input ajapjibi
\section{Translations in spacetimes\protect\\
with a straight cosmic string}
\input ajapstru
%
\section{Electromagnetic field}
\input ajapelmg

\end{document}

%% file: makro1.tex
\def \gab {g_{\a \b }}
\def \gmn {g_{\m \n }}

\def \gnn {g_{00}}
\def \gnj {g_{01}}
\def \gnd {g_{02}}
\def \gnt {g_{03}}

\def \gjj {g_{11}}
\def \gjd {g_{12}}
\def \gjt {g_{13}}

\def \gdd {g_{22}}
\def \gdt {g_{23}}

\def \gtt {g_{33}}
\def \mm {\mbox{\quad }}
\def \mv {\mbox{\qquad }}
\def \msip {\rightarrow}

\def \Fab {F_{\a \b }}
\def \Fmn {F_{\m \n }}

\def \Fnj {F_{01}}
\def \Fnd {F_{02}}
\def \Fnt {F_{03}}

\def \Fjd {F_{12}}
\def \Fjt {F_{13}}

\def \Fdt {F_{23}}

\def \Fumn {F^{\m \n }}

\def \Emn {E_{\m\n}}

\def \Tmn {T_{\m\n}}

%% file: makro2.tex
\def \a {\alpha}
\def \b {\beta}
\def \g {\gamma}

\def \d {\delta}

\def \ep {\epsilon}
\def \e {\eta}
\def \f {\phi}

\def \th {\theta}
\def \vth {\vartheta}

\def \l {\lambda}
\def \m {\mu}
\def \n {\nu}
\def \x {\xi}
\def \p {\pi}
\def \r {\rho}
\def \s {\sigma}
\def \t {\tau}

\def \o {\omega}
\def \z {\zeta}
\def \L {\pounds}

\def \der {\partial }
\def \nn {\nonumber}
\def \rov {\equiv}
\def \A {{\cal A}}
\def \BB {{\cal B}}
\def \C {{\cal C}}
\def \D {{\cal D}}
\def \E {{\cal E}}
\def \F {{\cal F}}
\def \GG {{\cal G}}
\def \K {{\cal K}}

\def \L {{\cal L}}

\def \I {{\cal I}}

\def \fk  {f^{(k)}}
\def \fmj {f^{(-1)}}
\def \fn  {f^{(0)}}
\def \fj  {f^{(1)}}

\def \Ak  {A^{(k)}}
\def \Amj {A^{(-1)}}
\def \An  {A^{(0)}}
\def \Aj  {A^{(1)}}

\def \gk  {g^{(k)}}
\def \gmj {g^{(-1)}}
\def \gen  {g^{(0)}}
\def \gej  {g^{(1)}}

\def \bu  {\b,_{u}}

\def \br  {\b,_{r}}

\def \bth {\b,_{\th}}

\def \bfi  {\b,_{\f}}

\def \brt   {\b,_{r\th}}

\def \brf   {\b,_{r\f}}

\def \bthth {\b,_{\th\th}}

\def \btf   {\b,_{\th\f}}

\def \bff   {\b,_{\f\f}}

\def \gu  {\g,_{u}}

\def \gr  {\g,_{r}}

\def \gth {\g,_{\th}}

\def \gf  {\g,_{\f}}

\def \grr   {\g,_{rr}}

\def \grt   {\g,_{r\th}}

\def \grf   {\g,_{r\f}}

\def \gthth {\g,_{\th\th}}

\def \gff   {\g,_{\f\f}}

\def \du  {\d,_{u}}

\def \dr  {\d,_{r}}

\def \dth {\d,_{\th}}

\def \df  {\d,_{\f}}

\def \drr   {\d,_{rr}}

\def \drt   {\d,_{r\th}}

\def \drf   {\d,_{r\f}}

\def \dthth {\d,_{\th\th}}

\def \dtf   {\d,_{\th\f}}

\def \dff   {\d,_{\f\f}}

\def \Vr   {V,_{r}}

\def \Vth  {V,_{\th}}

\def \Wr   {W,_{r}}

\def \Wth  {W,_{\th}}

\def \Wf   {W,_{\f}}

\def \Wrt   {W,_{r\th}}

\def \Wrf   {W,_{r\f}}

\def \Ur  {U,_{r}}

\def \Uth {U,_{\th}}

\def \Ufi  {U,_{\f}}

\def \Urt   {U,_{r\th}}

\def \Urf   {U,_{r\f}}

\def \sn {\sin \th}
\def \cs {\cos \th}
\def \tg {\tan \th}
\def \ctg {\cot \th}
\def \csec {\csc \th}
\def \dcsec {\csc^2 \th}
\def \ct {\bar t}
\def \dsn {\sin^2 \th}
\def \tsn {\sin^3 \th}

\def \dctg {\cot^2 \th}
\def \chd  {\cosh 2\d}
\def  \shd  {\sinh 2\d}
\def \mm {\mbox{\quad }}
\def \mv {\mbox{\qquad }}
\def \msip {\rightarrow}

\def \lkz  {\bigl(}
\def \pkz  {\bigr)}
\def \lvkz {\Bigl(}
\def \pvkz {\Bigr)}
\def \lvvkz {\biggl(}
\def \pvvkz {\biggr)}
\def \lhz  {\bigl[}
\def \phz  {\bigr]}
\def \lvhz {\Bigl[}
\def \pvhz {\Bigr]}

\def \pvsz {\Bigl\} }
\def \lvsz {\Bigr\{ }
\def \lvvsz {\Biggl\{}
\def \pvvsz {\Biggr\}}
\def \pul {\frac{1}{2}}
\def \B  {\frac{1}{2}B}
\def \V  {\frac{V}{r}}

\def \AnpB {\An+\B}
\def \AnmB {\An-\B}
\def \ABV {\lvkz A\emdb-\frac{1}{2r}BV\pvkz}
\def \ee {(e,_\th +e\ctg)}
\def \cc {(c,_\th+2c\ctg)}
\def \dd {(d,_\th+2d\ctg)}

\def \FUW  {(\Fnj-U\Fjd-W\Fjt\csec)}
\def \VUW {\lvkz\V\edb-r^2\edg U^2\chd
           -r^2\emdg W^2\chd-2r^2 UW\shd\pvkz}

\def \ffc  {(\fn-\fmj c)}

\def \edb  {e^{2\b }}
\def \emdb {e^{-2\b }}

\def \edg  {e^{2\g }}
\def \emdg {e^{-2\g }}
\def \eg   {e^{\g }}
\def \emg  {e^{-\g }}
\def \edbmg {e^{2(\b -\g)}}

\def \edbg  {e^{2(\b +\g)}}

\def \co    {(2ch^2-1)}
\def \Knn {K_{00}}
\def \Knj {K_{01}}
\def \Knd {K_{02}}
\def \Knt {K_{03}}
\def \Kjj {K_{11}}
\def \Kjd {K_{12}}
\def \Kjt {K_{13}}
\def \Kdd {K_{22}}
\def \Kdt {K_{23}}
\def \Ktt {K_{33}}

\def \GGjdt {G_{123}}
\def \GGnjd {G_{012}}
\def \GGnjt {G_{013}}
\def \GGndt {G_{023}}

\def \Jn {J^0}
\def \Jj {J^1}
\def \Jd {J^2}
\def \Jt {J^3}
\def \Rmn {R_{\m \n }}

\def \BE {\begin{equation}}
\def \EE {\end{equation}}
\def \BDM {\begin{displaymath}}
\def \EDM {\end{displaymath}}
\def \BEAH {\begin{eqnarray*}}
\def \EEAH {\end{eqnarray*}}
\def \BEA {\begin{eqnarray}}
\def \EEA {\end{eqnarray}}
\def \BM {\begin{math}}
\def \EM {\end{math}}

%% file: ajkap1.tex
\section{Introduction and Summary}
There is only one class of explicitly known radiative
solutions which are asymptotically flat in the~sense that
null infinity is global, i.e., admits spherical sections,
though its generators are not complete: boost-rotation symmetric
spacetimes describing "uniformly accelerated particles" of
various kinds \cite{bicakobecne}. A unique role of these
solutions is  exhibited by a theorem which roughly states
that in axially symmetric, asymptotically flat spacetimes
(in the~sense that at least a local null infinity exists)
the~only additional
symmetry that does not exclude radiation is the~boost symmetry
\cite{bicak}.

The~boost-rotation symmetric spacetimes have been used
in various contexts. From a mathematical point of view, these
solutions contain the~only known spacetimes in which
{\it{arbitrarily strong initial data}} with the~given symmetry
can be chosen on a hyperboloidal hypersurface which lead to
the~complete, smooth null infinity and regular timelike
infinity in future. From a more physical point of view these
are - and most probably will long remain - the~only exact solutions
of Einstein's equations for which one can find such quantities
as angular distribution of gravitational radiation emitted
by particles (represented by singularities) or by uniformly
accelerated black holes \cite{bicakzeleny}. In numerical
calculations these spacetimes have been employed as important
test beds in the~null cone version of numerical relativity
\cite{wini,gomez}, and, most recently, also in the~standard
approach based on a spacelike initial hypersurface \cite{alcu}.
In the~context of quantum gravity the~boost-rotation symmetric
spacetimes (as "generalized C-metric") have been used to
describe production of black-hole pairs in strong background fields
(see e.g. \cite{hawking}). We refer to the~recent review \cite{bicakLes}
of exact radiative spacetimes and to the~comprehensive analysis
of the~general structure of the~boost-rotation symmetric spacetimes
\cite{bicakobecne} for more details and references on both
the~history and recent developments in these issues.

Until now all work on the~general properties of the~boost-rotation
symmetric spacetimes has concentrated on the~vacuum case with two
non-null, hypersurface orthogonal Killing vectors - the~rotational
(axial) Killing vector and the~boost Killing vector. The~metric
of such spacetimes, in suitable coordinates, contains just
two functions. It can be constructed starting from
the~appropriately behaved boost-rotation symmetric
solution of the~flat-space wave equation with sources
which is satisfied by one of the~functions. The~other
metric function can then be determined by a quadrature
(see \cite{bicakobecne}). In the~present paper we start
the~systematic study of more general boost-rotation symmetric
spacetimes. We assume that an {\it{electromagnetic field}} coupled
to gravity may also be present; and we consider {\it{Killing vectors}}
which need {\it{not}} be {\it{hypersurface orthogonal}}. Even if
only one of these extensions is taken into account, the~field
equations become fully nonlinear - there is no flat-space wave
equation available. The~inclusion of electromagnetic field is
also of interest because there exist the~analogous features of
gravitational fields due to uniformly accelerated masses in
general relativity and of electromagnetic fields due to uniformly
accelerated charges in special relativity. The~study of
boost-rotation symmetric fields within the~Einstein-Maxwell theory
treats both gravitational and electromagnetic fields from
a unified point of view.

It is known that the~example of a~boost-rotation symmetric
electrovacuum solution with Killing vectors which are not
hypersurface orthogonal exists - this is the~charged "rotating"
C-metric \cite{pleb}. Here we consider general solutions.
Our main result is the~theorem which roughly states that even
under the~presence of electromagnetic field and "rotating sources",
the~boost symmetry is the~only one which can be combined
with axial symmetry and radiation  exists. However we shall
also obtain other results (e.g. for the~news function and
total mass) some of which are new even in the~vacuum case
with hypersurface orthogonal Killing vectors.

In the~following Section II we start out from the~general form
of axially symmetric metric and electromagnetic field
in Bondi-Sachs coordinates $\{u,\ r,\ \th,\ \f\}$, where
$u=$ const labels null hypersurfaces (in flat space $u=t-r$ is
the~usual retarded time), $r$ is a luminosity distance along
null rays $u=$ const, $\th=$ const, $\f=$ const, and $\th$, $\f$
are standard spherical angles. We consider their asymptotic
expansions at $r\ \msip\ \infty$ as they follow from
the~Einstein-Maxwell equations under the~assumption of asymptotic
flatness. Since we wish to study spacetimes in which the~axial
Killing vector is in general not hypersurface orthogonal we cannot
employ the~original Bondi's et al work \cite{bondi} but have to
start from its generalization by Sachs \cite{bondi} or, rather,
from van der Burg's \cite{burg} explicit treatment of the~asymptotic
behaviour of the~coupled Einstein-Maxwell fields
in the~Bondi-Sachs coordinates. Now  van~der~Burg's
work involves complicated equations in which many errors and
misprints appear. These do not change basic conclusions of his
paper but we need correct forms. Therefore, in Appendix A we
first summarize all Einstein-Maxwell's equations in the~Bondi-Sachs
coordinates in general spacetime and then give
asymptotic forms of their solutions under the~assumption of axial
symmetry. (All equations were checked by using {\sc{MAPLE}} V.)
The~structure of the~field equations (the~splitting into twelve main
equations and five supplementary conditions), the~total quantities
(the~mass and the~charge) and their fluxes given in terms of two
gravitational and two electromagnetic news functions are briefly
reviewed also in Appendix A. The~resulting formulas are used
extensively in Section II to prove the~theorem which
states the~following:

Suppose that an axially symmetric electrovacuum spacetime admits a
"piece" of null infinity. If this spacetime admits an additional
Killing vector forming with the~axial Killing vector a
$2$-dimensional Lie algebra, then the~additional Killing vector
either generates a supertranslation or it is the~boost Killing
vector.

The~theorem is proven by first decomposing the~additional Killing
vector field $\e^\a$ in the~null (Sachs) tetrad and then solving
the~Killing equations asymptotically in the~leading terms. In
order to illustrate the~form of the~Killing equations, the~lengthiest
among them, $\L_\e \gnn=0$, is written down in Appendix~B.

In Section III the~case of the~supertranslational Killing field is
considered further. By solving Killing equations in the~higher
orders in $r^{-k}$ and  considering also asymptotic
solutions of  equations $\L_\e \Fmn =0$
(assuming thus that the~electromagnetic field shares the~same
symmetry),
we show that the~supertranslational Killing
field has in fact to be the~generator either of  translations along
the~axis of axial symmetry ($z$-axis) or time translations. Resulting
spacetimes are non-radiative. Somewhat surprisingly
perhaps, the~news functions of the~system need not necessarily be
vanishing. This is not because of cylindrical waves which of
course are symmetric under translations along $z$-axis.
Cylindrical waves are excluded from our analysis
if we
assume that the~local null infinity exists at $\th=\p /2$;
there are no cylindrical spacetimes which admit a regular
cross-section of null infinity. (We refer
to the~papers \cite{ABS1,ABS2} for the~study of null
infinity of cylindrically symmetric spacetimes and to our
forthcoming work in which the~Killing equations will be considered
in a greater detail in further orders in $r^{-k}$ for
translational Killing vectors.)
%(in the~present
%context see in particular the~remark at the~end of Section III,
%part B in Ref. \cite{ABS2}). Indeed,
Here we consider spacetimes with a straight cosmic string 
along $z$-axis which,
as shown by Bi\v c\' ak and Schmidt \cite{Bstruna},
%a flat spacetime with a deficit angle
%corresponding to an infinite cosmic string in Bondi's coordinates
have a non-vanishing news function independent of time. If we assume
that electromagnetic field is regular  at the~axis of symmetry,
only the~gravitational news function is non-vanishing. We give
the~explicit form of the~translational Killing vector in this case.
This translational Killing vector in the~case of the~string
has not been given in Ref. \cite{bicak} because of
a sign error. The~error and misprints appearing in  \cite{bicak}
are corrected in Appendix C. In Appendix D  the~translations
in Bondi's coordinates in a spacetime with a
straight, non-rotating cosmic string are analyzed in detail. At
the~end of Section III we give explicitly the~asymptotic form of
both gravitational and electromagnetic fields in the~Bondi-Sachs
coordinates for stationary spacetimes without a string.

Section IV is devoted to the~case when the~additional Killing
vector is the~boost Killing vector. By analyzing the~Killing
equations  in further orders in $r^{-k}$
and considering also  equations $\L_\e \Fmn=0$,
we discover that both
gravitational and electromagnetic news functions must have
the~same functional form - they depend on $u$ and $\th$ so that
\mbox{the~news function\ } \mbox{$=f(\sn /u)u^{-2}$}, where $f$ is an
arbitrary function of its argument. We also derive the~general
form of the~mass aspect and total (Bondi) mass for boost-rotation
symmetric spacetimes. The~mass aspect is given by
\BE
M(u,\th)=\frac{1}{2\sn}(w^2\K,_w),_w+\frac{\L}{u^3}\ ,\nn
\EE
where $w=\sn /u$, $\K(w)$ is an arbitrary function, and
$\L(w)=\l(w) /w^3$, where $\l(w)$ satisfies the~simple equation
(\ref{boostlambda}) in the~main text. $\l(w)$ can be determined if
the~news functions are known. The~total Bondi mass is then given
by
\BE
m(u) =\frac{1}{4}\int_0^\p (w^2\K,_w),_wd\th
     +\frac{1}{2}\int_0^\p \frac{\L w}{u^2}d\th\ .\nn
\EE
The~formulas for the~mass aspect and total mass are new even in
the~case of hypersurface orthogonal  Killing vectors studied in
\cite{bicak}. Also, the~boost Killing vector is here expanded to
further orders than in Ref. \cite{bicak} (see Eq. (\ref{etaboost})).

Let us note that in general Bondi's news function has an invariant
meaning only if null infinity is topologically
$S^2\times {\rm R}$. However, as shown in \cite{Bstruna},
in axially symmetric spacetime relative to a chosen
$\der /\der\f$ Killing vector Bondi news and translations
have an invariant meaning even locally.

Our conventions for the~Riemann and Ricci tensors follow those
of Ref. \cite{MTW} but our signature is $-2$.

%% file: ajkap2.tex
\section{Axially symmetric space-times with another symmetry}

Consider an axially symmetric electrovacuum spacetime with
circular group orbits; denote the~corresponding Killing vector
field by $\der / \der\f$. Assume that at least the~"piece of
$\I^+$" exists in the~sense of \cite{AstSchm}. Then one can
introduce the~Bondi-Sachs coordinate system
\mbox{ \{ $u$,~$r$,~$\th$,~$\f$ \} $\rov$ \{
             $x^0$,~$x^1$,~$x^2$,~$x^3$\}}
in which the~metric satisfying the~Einstein-Maxwell equations
has the~form
\cite{bondi,burg}
\BEA
ds^2&=&\VUW du^2\nn\\
    & &\ + 2\edb du dr
              +2r^2(\edg U\chd +W\shd)  dud\th
              +2r^2(\emdg W\chd +U\shd)\sn\ dud\f\label{ds}\\
    & &\ - r^2\left[ \chd(\edg d\th^2 +\emdg \dsn\ d\f^2 )
                        +2\shd\sn\ d\th d\f\right] \ ,\nn
\EEA
where six metric functions $U$, $V$, $W$, $\b$, $\g$, $\d$ and
the~Maxwell field $\Fmn$ do not depend on $\f$ because of axial
symmetry.

The~complete set of the~Einstein-Maxwell equations and  the~asymptotic
expansions of the~metric functions and electromagnetic field
components at large $r$ are given in Appendix A. We shall need
these expansions, in which
also the~field equations are used,
up to the~following orders (all "coefficients" $c$, $d$, $M$, $e$,
$f$, ... below being in general functions of $u$ and $\th$)
\BEA
\g  &=&\frac{c}{r}+O(r^{-3})\ ,\nn\\
\d  &=&\frac{d}{r}+O(r^{-3})\ ,\nn\\
\b  &= & -\frac{1}{4}(c^2+d^2)\frac{1}{r^2}+O(r^{-4})\ ,\nn\\
U   &= &-\cc\frac{1}{r^2}+O(r^{-3})\ ,\label{rozvmetr}\\
W   &= &-\dd\frac{1}{r^2}+O(r^{-3})\ ,\nn\\
V   &= &r-2M+O(r^{-1})\ .\nn
\EEA
The~leading terms in the~electromagnetic field tensor are given
by
\BEA
\Fjd&=&\frac{e}{r^2}+(2E+ec+fd)\frac{1}{r^3}+O(r^{-4})\ ,\nn\\
\Fjt&=&\lvkz\frac{f}{r^2}+(2F+ed-fc)\frac{1}{r^3}
              +O(r^{-4})\pvkz\sn\ ,\nn\\
\Fnj&=&-\frac{\ep}{r^2}+(e,_\th+e\ctg)\frac{1}{r^3}
              +O(r^{-4})\ ,\label{rozvpole}\\
\Fdt&=&\lvkz -\m-(f,_\th+f\ctg)\frac{1}{r}+O(r^{-2})\pvkz\sn\ ,\nn\\
\Fnd&=&X+(\ep,_\th-e,_u)\frac{1}{r}+O(r^{-2})\ ,\nn\\
\Fnt&=&\lvkz Y-\frac{f,_u}{r}+O(r^{-2})\pvkz\sn\ .\nn
\EEA
The~mass aspect $M(u,\th)$ is connected with the~two gravitational
news functions $c,_u$ and $d,_u$, and with the~electromagnetic news
 functions $X$ and  $Y$ by the~relation
\BE
M,_u=-(c,_u^2+d,_u^2)-(X^2+Y^2)
     +\pul (c,_{\th\th}+3c,_\th\ctg-2c),_u\ .\label{Mu}
\EE
As a consequence of the~Einstein-Maxwell equations
the~electromagnetic functions $e$, $f$, $\ep$, $\m$ and $X$, $Y$ are
connected by the~relations
\BEA
e,_u  &=&\frac{1}{2}\ep,_\th-(cX+dY)\ ,\label{udere}\\
f,_u  &=&-\frac{1}{2}\m,_\th-(-cY+dX)\ ,\label{uderf}\\
\ep,_u&=&-X,_\th-X\ctg\ ,\label{uderep}\\
\m,_u &=&-Y,_\th-Y\ctg\ .\label{udermi}
\EEA
The~gravitational news functions enter the~evolution equations
for $e$ and $f$. The~evolution equations for functions $E$, $F$
are given in Appendix A. There  the~evolution of the~whole
Einstein-Maxwell system is summarized and the~interpretation of
the~mass and the~charges of the~system is recalled. Here let us
only notice that both the~gravitational news functions $c,_u$,
$d,_u$ and the~electromagnetic news functions $X$, $Y$ are freely
specifiable  functions.

Now we wish to emphasize that we assume that Eqs.
(\ref{ds})~-~(\ref{udermi}) are valid for all
$\f\in[0,2\p)$, however, they need {\it {not}}
to be true for all $\th\in[0,\p]$. Since we want to admit
spacetimes with only "local" $\I^+$, we assume Eqs.
(\ref{ds})~-~(\ref{udermi}) to be satisfied in some open interval
of $\th$, i.e., not necessarily on the~whole sphere. In
particular, the~"axis of symmetry" ($\th=0,\p$) may contain some
nodal singularities and  need not thus  be regular. The~regularity
conditions on the~axis (which can be obtained by transforming
(\ref{ds}) into local Minkowskian coordinates) on functions
$\g$, $V$, $\b$ ... need not be satisfied for any $u$. If the~axis
is singular then at least two generators of $\I^+$ would be
missing so that $\I^+$ would not be topologically
$S^2\times {\rm R}$ (this is exactly the~case which can occur in
the~boost-rotation symmetric spacetimes discussed in Sec. IV).

Let us now assume that another Killing vector field $\e$ exists
which forms together with $\x =\der / \der\f$ a two-parameter
group. In Ref. \cite{bicak}
it is  proved (see Lemma in Sec. 2) that in the~case of $\x$ with
circles as integral curves, $\x$ and $\e$ determine an abelian
Lie algebra so that we can assume $[\e,\x]=0$. Hence,
the~components of $\e^\a$ are independent of $\f$.

Introduce the~standard null tetrad
field  \cite{sachs}
$\lbrace k^{\a }, m^{\a }, t^{\a }, \ct ^{\a }\rbrace$, with bar
denoting the~complex conjugation, where  $k_\a=\der_\a u$,
$m^\a k_\a =1$, $m^\a m_\a =0$, and the~complex vector
$t^\a=t_R^\a+it_I^\a$ (subscripts $R$ and $I$ denoting the~real
and imaginary parts) obeys
$t^\a \ct _\a =-1$, $t^\a t_\a = t^\a k_\a = t^\a m_\a = 0$.
A convenient choice of the~tetrad reads
\BEA
k_\a&=&\lvhz 1\ ,\ 0\ ,\ 0\ ,\ 0\pvhz\ ,\mm
m_\a=\lvhz\pul Vr^{-1}\edb\ ,\ \edb\ ,\ 0\ ,\ 0\pvhz\ ,\nn\\
t_\a&=&\pul r(\chd)^{-\pul}
        \lvhz (1+\shd)\eg U+\chd\emg W
           +i[(1-\shd)\eg U-\chd\emg W]\ ,\label{Btetrada}\\
    &\ &\mm 0\ ,\
         -(1+\shd+i(1-\shd))\eg\ ,\ -(1-i)\chd\sn\emg\pvhz\ .\nn
\EEA
This choice indeed implies
$\gmn = 2 k_{(\m } m_{\n )} - 2 \ct_{(\m } t_{\n )}$, with $\gmn$
given by (\ref{ds}).

We now decompose the~Killing vector field $\e^\a$ in the~null
tetrad,
\BE
\e ^\a  = A k^\a +B  m^\a +C t^\a +\bar C\ct^\a\ , \label{bbeta}
\EE
or
\BE
\e ^\a  =Ak^\a +Bm^\a +\tilde{f}(t_{R}^{\a}+t_{I}^{\a})
                      +\tilde{g}(t_{R}^{\a}-t_{I}^{\a})\ ,
\label{Beta}
\EE
where $A$, $B$, $C=C_R +iC_I$, $\tilde{f}=C_R-C_I,\
\tilde{g}=C_R  +C_I $ are general functions of $u$, $r$, $\th$.
The~Killing equations
\BE
\L_\e \gab=0\label{Kill}
\EE
can then be rewritten in the~form
\BEA
0=\L_\e\gab &=&A\L_k\gab +B\L _m\gab
             +2A,_{(\a }k_{\b )}+2B,_{(\a } m_{\b )}\nn\\
            & &+\tilde{f} \lbrack (\L _t \gab)_R
                               + (\L _t \gab)_I \rbrack
             +2\tilde{f},_{(\a } \lbrack t_{R\b )} + t_{I\b )}\rbrack \\
             & &
        +\tilde{g} \lbrack (\L _t \gab )_R - (\L _t \gab )_I \rbrack +
              2 \tilde{g},_{(\a } \lbrack t_{R\b )} - t_{I\b )}\rbrack
                       \ . \nn
\EEA
The~easiest is the~equation
\BE
\L_\e \gjj=2\edb B,_r=0 \ ,\nn
\EE
which implies
\BE
B=B(u,\th)\ .\label{B}
\EE
In the~hypersurface orthogonal case, analyzed in Ref. \cite{bicak},
also the~equations
$\L_\e \gnt=\L_\e \gjt=\L_\e \gdt=0$ can easily be
solved. This is not so if $\der /\der\f$ is not hypersurface
orthogonal. All Killing equations other than $\L_\e \gjj=0$
become now complicated. For illustration, equation
$\L_\e \gnn=0$ is written down in Appendix~B; the~other
equations can be found in \cite{aja}.

We assume that the~coefficients $A$, $\tilde{f}$, $\tilde{g}$ can be expanded in
powers of $r^{-k}$ and solve the~Killing equations
asymptotically. Denoting by $\Ak$, $\fk$, $\gk$ the~coefficients
of $r^{-k}$ in the~expansions, we first notice that
$\L_\e \gdd=0$, \mbox{$\L_\e \gjd=0$}, $\L_\e \gjt=0$ imply
\BEA
A&=&\Amj r+\An+\frac{\Aj}{r}+O(r^{-2})\ ,\nn\\
\tilde{f}&=&\fmj r+\fn+\frac{\fj}{r}+O(r^{-2})\ ,\label{Afg}\\
\tilde{g}&=&\gmj r+\gen+\frac{\gej}{r}+O(r^{-2})\ ,\nn
\EEA
where $\Ak$, $\fk$, $\gk$ are functions of $u$ and $\th$.
Remaining nine Killing equations imply
the~conditions on  functions
$B$, $\Ak$, $\fk$, $\gk$.
The~procedure of their solutions is similar to that
in Ref. \cite{bicak}
(ef. Eqs. (20)-(24) therein). However, the~equations are now more
complicated and there are three additional equations to be
satisfied.

In the~leading orders in $r^{-k}$ the~Killing equations can be
written down easily (we omit equations
$\L_\e \gjd=0$, $\L_\e \gjt=0$ since they do not restrict
the~leading terms in $\e^\a$):
\BEA
\ \L_\e \gnn
              &=&0 \mm(r^{\ 1}):\ \Amj,_u=0\ ,\label{BKjj}\\
\ \L_\e \gnj
              &=&0 \mm(r^{\ 0}):\ B,_u+\Amj=0\ ,\label{BKdn}\\
\ \L_\e \gnd
              &=&0 \mm(r^{\ 2}):\ \fmj,_u=0\ ,\label{BKtd}\\
\ \L_\e \gnt
              &=&0 \mm(r^{\ 2}):\ \gmj,_u=0\ ,\label{BKcd}\\
\ \L_\e \gdd
              &=&0 \mm(r^{\ 2}):\ \fmj,_\th+\Amj=0\ ,\label{BKod}\\
\ \L_\e \gdt
              &=&0 \mm(r^{\ 2}):\ -\gmj,_\th+\gmj\ctg=0
                           \ ,\label{BKded}\\
\ \L_\e \gtt
              &=&0 \mm(r^{\ 2}):\ \fmj\ctg+\Amj=0\ .\label{BKdsd}
\EEA
The~system of equations
(\ref{BKjj}) - (\ref{BKtd}), (\ref{BKod}),
(\ref{BKdsd})
is, at this order, identical to Eqs. (20) - (24) in
Ref. \cite{bicak} . The~solutions are thus identical, reading
\BEA
\Amj&=&k\cs\ ,\nn\\
\fmj&=&-k\sn\ ,\label{BAmjfmjB}\\
B   &=&-ku\cs+\a(\th)\ ,\nn
\EEA
where $k=\mbox{const}$ and $\a$ is an arbitrary function of $\th$. Eqs.
(\ref{BKcd}) and (\ref{BKded})
imply
\BE
\gmj=h\sn\ ,\label{Bgmj}
\EE
where $h=\mbox{const}$.

Regarding Eqs. (\ref{Btetrada}), (\ref{Beta}) and (\ref{Afg}) we
find $\e^\f=h+O(r^{-1})$. Since the~contribution of $h$ to the~vector
field $\e^\a$ is just $\e^\f=\mbox{const}$, which is a constant
multiple of the~axial Killing vector  $\der /\der\f$, we may,
without loss of generality, put $h=0$.
Therefore, in the~lowest order in $r^{-1}$ the~general
asymptotic form of the~Killing vector $\e$ turns out to
be
\BE
\e^\a=[-ku\cs+\a(\th)\ ,\
          kr\cs+O(r^{0})\ ,\ -k\sn+O(r^{-1})\ ,\ O(r^{-1})]\ , \label{Bbotr}
\EE
where $k$ is a constant, $\a$ -- an arbitrary function of $\th$.
Thus, in
the~leading order of the~asymptotic expansion the~presence of
electromagnetic field satisfying the~boundary conditions
(\ref{rozvpole})
and the~fact that the~axial Killing vector need not be
hypersurface orthogonal do not change the~conclusion obtained
in Ref. \cite{bicak}
in the~vacuum case with hypersurface orthogonal $\der /\der\f$.
When $k=0$, the~vector field
(\ref{Bbotr})
generates
supertranslations. We shall see in the~next section that
the~resulting spacetimes are then non-radiative.

Assuming $k\not= 0$,
it is then easy to find
a Bondi-Sachs coordinate system with $\a=0$ by making
a supertranslation  $\tilde{u} =u+\hat{\a}(\th)$, $\hat{r}=r$,
$\hat{\th}=\th$, $\hat{\f}=\f$, where $\hat{\a}$ satisfies
the~equation $-\hat{\a},_\th k\sn+\hat{\a} k\cs=-\a$
(cf. \cite{bicak}). (Notice that no singularity arises for
$\hat{\a}$ even if $\a (0)\not= 0$.)
We then obtain $\e^{\tilde{u}}=-k\tilde{u}\cs$.
Hence, we put $\a=0$ in Eq. (\ref{Bbotr}) and without loss
of  generality (buno) we choose $k=1$. Then $B=-u\cs$,
$\Amj=\cs$, $\fmj=-\sn$ and $\gmj=0$.
The asymptotic form of the~Killing vector field
$\e$ is 
\BE
\e^\a=[-u\cs\ ,\ r\cs+O(r^{0})\ ,
\ -\sn+O(r^{-1})\ ,\ O(r^{-1})]\ ,\label{vboost}
\EE
which is the~{\it{boost Killing vector}}. It  generates
the~Lorentz transformations along the~axis of axial symmetry.

We have thus proven the~following

{\bf Theorem}:\ Suppose that an axially symmetric
electrovacuum spacetime admits a "piece " of $\I^+$ in the~sense
that the~Bondi-Sachs coordinates can be introduced in which
the~metric takes the~form (\ref{ds}) - (\ref{rozvmetr}) and
the~asymptotic form of the~electromagnetic field is given by
(\ref{rozvpole}). If this spacetime admits an additional Killing
vector forming with the~axial Killing vector a 2-dimensional Lie
algebra, then the~additional Killing vector has asymptotically
the~form (\ref{Bbotr}). For $k=0$ it generates a
supertranslation; for $k\not= 0$ it is the~boost Killing field.

On $\I^+$ in the~coordinates $u$, $l=r^{-1}$, $\th$, and $\f$
the~boost Killing vector reads (cf. \cite{bicak})
\BE
\e^\a_{/\I^+}=[-u\cs\ ,\ 0\ ,\ -\sn\ ,\ 0]\ .
\EE

%% file: ajkap3.tex
\section{ The~supertranslational Killing field}
In this section we shall show that if the~Killing field
(\ref{Bbotr})
for $k=0$ is the~supertranslational Killing field, it has, in
fact, to be the~generator of translations
and the~resulting spacetime is thus non-radiative.

Assuming $k=0$, and considering the~Killing equations (\ref{Kill})
in the~higher orders in $r^{-k}$ than in Eqs. 
(\ref{BKjj}) - (\ref{BKdsd}),
and taking into account the~metric
expansions (\ref{rozvmetr}), we find the~following restrictions
on the~expansion coefficients of functions
$A$, $\tilde{f}$ and $\tilde{g}$:
\BEA
\ \L_\e \gnn  &=&0\mm(r^{\ 0}):\ \An,_u=0\ ,\label{tBKjn}\\
              & &\ \mm(r^{-1 }):\ \Aj,_u
                                 -\fn,_u\cc -\gen,_u\dd=0\ ,
                                     \label{tBKjmj}\\
\ \L_\e \gnj  &=&0\mm(r^{-1 }):\ \mbox{satisfied automatically}\ ,
                                   \label{tBKdmj}\\
              & &\ \mm(r^{-2 }):\ -B(cc,_u+dd,_u)-BM-\Aj
                              +\fn\cc+\gen\dd=0\ ,\label{tBKdmd}\\
\ \L_\e \gnd  &=&0\mm(r^{\ 1}):\ \fn,_u=0\ ,\label{tBKtj}\\
              & &\ \mm(r^{\ 0}):\    \lvkz\AnpB\pvkz,_\th
                                    +\fn c,_u-c\fn,_u
                             -\fj,_u -2\gen,_u d=0\ ,\label{tBKtn}\\
\ \L_\e \gnt  &=&0\mm(r^{\ 1}):\ \gen,_u=0\ ,\label{tBKcj}\\
              & &\ \mm(r^{\ 0}):\ -\gen c,_u+\gen,_u c
                              -\gej,_u+2\fn d,_u=0\ ,\label{tBKcn}\\
\ \L_\e \gjd  &=&0\mm(r^{\ 0}):\ \fn +B,_\th=0\ ,\label{tBKssn}\\
              & &\ \mm(r^{-1 }):\ \fj-B\cc+\gen d=0 \ ,\label{tBKssmj}\\
\ \L_\e \gjt  &=&0 \mm(r^{\ 0}):\ \gen=0\ ,\label{tBKsen}\\
              & &\ \mm(r^{\ -1}):\ \gej-B\dd-\fn d=0
                                      \ ,\label{tBKsemj}\\
\ \L_\e \gdd  &=&0\mm(r^{\ 1}):\ \fn,_\th+Bc,_u+\AnmB=0
                          \ ,\label{tBKoj}\\
              & &\ \mm(r^{\ 0}):\ \fj,_\th+B(cc,_u+2dd,_u)
                                 -B\cc,_\th+\Aj+BM+
                          2d(\gen,_\th-\gen\ctg)=0\ ,\label{tBKon}\\
\ \L_\e \gdt  &=&0\mm(r^{\ 1}):\ 2Bd,_u+\gen,_\th-\gen\ctg
                              =0 \ ,\label{tBKdej}\\
              & &\ \mm(r^{\ 0}):\ -2\An d-4\fn d\ctg+B\dd,_\th
                                   -B\dd\ctg+Bd\nn\\
              & &\ \mm\mv\mv +\gen,_\th c-\gen(c,_\th+c\ctg)
                               -\gej,_\th +\gej\ctg=0\ ,
                           \label{tBKden}\\
\ \L_\e \gtt  &=&0\mm(r^{\ 1}):\ \fn\ctg-Bc,_u+\AnmB=0\ ,
                                   \label{tBKdsj}\\
              & &\ \mm(r^{\ 0}):\ \ctg\lvhz-B\cc+\fj-\fn c\pvhz
                                 -c,_\th\fn\nn\\
              & &\ \mm\mv\mv     +c(\AnmB)+\Aj+2Bdd,_u+BM=0\ .
                                \label{tBKdsn}
\EEA
First notice that Eqs. (\ref{tBKjn}), (\ref{tBKtj}),
(\ref{tBKcj}), (\ref{tBKssn}), (\ref{tBKsen}) and (\ref{tBKdej})
imply
\BE
\An=\An(\th)\ ,\ \fn=\fn(\th)=-B,_\th\ ,\
\gen=0\ ,\ d=d(\th)\ .\label{tAnfngnd}
\EE
Using these results in the~remaining equations we find
\BEA
\An &=&\pul(B,_{\th\th}+B,_\th\ctg+B)\ ,\label{tAn}\\
c,_u&=&\frac{1}{2B}(B,_{\th\th}-B,_\th\ctg)\ ,\label{tcu}\\
c&=&\frac{u}{2B}(B,_{\th\th}-B,_\th\ctg)+\o(\th)\ ,\label{tc}\\
\Aj &=&\Aj(\th)\ ,\label{tAj}\\
\fj &=&B\cc\ ,\label{tfj}\\
\gej&=&B\dd-B,_\th d\ ,\label{tgj}
\EEA
where $\o(\th)$ is an arbitrary function of $\th$; $\o$ can be
transformed away by a supertranslation $u=\bar{u}+\hat{\a}(\th)$ with
$\hat{\a}$ satisfying $\o +(-\hat{\a},_{\th\th}+\hat{\a},_{\th}\ctg )/2=0$.
Equations (\ref{tBKtn}) and (\ref{tBKden}) are now satisfied
identically.
Equations (\ref{tBKdmd}), (\ref{tBKon}) and (\ref{tBKdsn}) lead
to only one further independent condition:
\BE
M=-cc,_u-B^{-1}[\Aj+B,_\th\cc]\ .\label{tM}
\EE
Important results follow from Eqs. (\ref{tAnfngnd}) and
(\ref{tcu}). Eq. (\ref{tAnfngnd}) shows that the~news function
$d,_u=0$. Since $B=B(\th)$, Eq. (\ref{tcu}) implies that
the~time-derivative of the~news function $c,_u$ must vanish.
Therefore, with the~supertranslational Killing field, we arrive
at the~Weyl tensor (see e.g. \cite{burg})
\BE
C_{\a\b\g\d}m^\a t^\b m^\g t^\d
              =[(c+id)_{,uu}]\frac{1}{r}+O(r^{-2})\ ,
         \label{Weyl}
\EE
which is non-radiative since the~first
term proportional to $r^{-1}$ vanishes.

Substituting the~expansion of the~metric functions (\ref{rozvmetr})
into the~null tetrad  (\ref{Btetrada}) and
coefficients $A$, $B$, $\tilde{f}$ and $\tilde{g}$ (given by
Eqs. (\ref{tAnfngnd})-(\ref{tM}))
into Eq. (\ref{Beta}), we find the~expansion
of the~supertranslational Killing vector to be
\BEA
\e^\m&=&\lvhz B(\th)\ ,\ \pul (B,_{\th\th}+B,_\th\ctg)+
               [-B,_{\th\th}-2B,_\th B,_{\th\th\th}
             +2{B,_\th}^2B,_{\th\th}B^{-1}-2{B,_\th}^3\ctg B^{-1}
                 \label{etasupertrans}\\
      &\ &\mm +{B,_\th}^2 (3\ctg^2-\frac{2}{\dsn})]B^{-1}\frac{u}{4r}
                    +O(r^{-2})\ ,
       \ -B,_\th \frac{1}{r}+B,_\th\frac{c}{r^2}+O(r^{-3})\ ,
       \  B,_\th \frac{d}{r^2}\sn+O(r^{-3})\pvhz\ .\nn
\EEA

Let us now turn to the~asymptotic properties of electromagnetic field.
We assume $\L_\e \Fab=0$ where $\e$ is now
the~supertranslational Killing field (\ref{etasupertrans}).
From the~general expression of the~Lie derivative of $\Fmn$
with respect to general $\e^\a$ decomposed into the~null tetrad
according to Eq. (\ref{Beta}),
\BEA
\L_\e \Fab&=&A\L_k \Fab
               +A,_\a k^\g F_{\g\b}+A,_\b k^\g F_{\a\g}\nn\\
          &\ &    +B\L_m \Fab
                +B,_\a m^\g F_{\g\b}+B,_\b m^\g F_{\a\g}\nn\\
          &\ & +\tilde{f}[(\L_t \Fab)_R+(\L_t \Fab)_I]
               +\tilde{g}[(\L_t \Fab)_R-(\L_t \Fab)_I]\label{E1}\\
          &\ & +\tilde{f},_\a (t^\g_R+t^\g_I) F_{\g\b}
               +\tilde{f},_\b (t^\g_R+t^\g_I) F_{\a\g}\nn\\
          &\ & +\tilde{g},_\a (t^\g_R-t^\g_I) F_{\g\b}
               +\tilde{g},_\b (t^\g_R-t^\g_I) F_{\a\g}\ ,\nn
\EEA
and after substituting for $A$, $B$, $\tilde{f}$, and $\tilde{g}$ in accordance
with Eq. (\ref{etasupertrans}), we find that
in the~first orders in $r^{-k}$ the~Lie equations imply
\BEA
\L_\e \Fnj&=&0\mm (r^{- 2}):\mm -\ep,_u B+XB,_\th=0\ ,\label{E3}\\
\L_\e \Fnd&=&0\mm (r^{\ 0}):\mm X,_u B=0\ ,\label{E4}\\
\L_\e \Fnt&=&0\mm (r^{\ 0}):\mm (Y\sn),_u B=0\ ,\label{E5}\\
\L_\e \Fjd&=&0\mm (r^{- 2}):\mm e,_u B+\ep B,_\th =0\ ,\label{E6}\\
\L_\e \Fjt&=&0\mm (r^{- 2}):\mm (f\sn),_u B-\m\sn B,_\th=0\ ,\label{E7}\\
\L_\e \Fdt&=&0\mm (r^{\ 0}):\mm -(\m\sn),_u B+Y\sn B,_\th =0\ .\label{E8}
\EEA
From Eq. (\ref{E4}) we get $X=X(\th)$. Combining Eq. (\ref{E3})
and Maxwell equation (\ref{uderep}) we get equation for $X$,
$(XB\sn),_\th=0$, the~solution being
\BE
X=\frac{x_0}{B\sn}\ ,\ x_0=\mbox{const}\ .\label{E9}
\EE
Regarding the~last result and integrating Eq. (\ref{E3}), we obtain
\BE
\ep=\frac{x_0B,_\th}{B^2\sn}u+\ep_1(\th)\ ,\label{E10}
\EE
where $\ep_1(\th)$ is an arbitrary integration function.
Eq. (\ref{E6}) then implies
\BE
e=-\frac{x_0B,_\th^2}{2B^3\sn}u^2-\frac{\ep_1 B,_\th}{B}u
+ e_1(\th)\ ,\label{E11}
\EE
where $e_1(\th)$ is again an integration function.

Analogously, using Eqs. (\ref{E5}), (\ref{E7}), (\ref{E8})
and Maxwell equation (\ref{udermi}) we find
\BEA
Y&=&\frac{y_0}{B\sn}\ ,\ y_0=\mbox{const}\ ,\label{E13}\\
\m&=&\frac{y_0 B,_\th}{B^2\sn}u+\m_1 (\th)\ ,\label{E14}\\
f&=&\frac{y_0 B,_\th^2}{2B^3\sn}u^2+\frac{\m_1 B,_\th}{B}u+
          f_1 (\th)\ ,\label{E15}
\EEA
where $\m_1$, $f_1$ are arbitrary functions of $\th$.

Comparing Eqs. (\ref{E11}) and (\ref{E15}) with
Maxwell equations (\ref{udere}) and (\ref{uderf})
we obtain restrictions on $\ep_1$ and $\m_1$:
\BEA
{\ep_1},_\th+\frac{2\ep_1 B,_\th}{B}%-\frac{2x_0\o(\th)}{B\sn}
                 -\frac{2y_0 d}{B\sn}&=&0\ , \label{E12}\\
{\m_1},_\th+\frac{2\m_1 B,_\th}{B}%-\frac{2y_0\o(\th)}{B\sn}
                 +\frac{2x_0 d}{B\sn}&=&0\ . \label{E16}
\EEA

Now we assume that the~Killing field $\e$
is bounded at the~axis, i.e. $B(\th)$ is bounded for
$\th=0,\p$ and the~electromagnetic field (\ref{rozvpole}), i.e.
$X$ and $Y$ have to be bounded there, too. Therefore
constants entering $X$ and $Y$ must vanish:
\BE
x_0=y_0=0\ .
\EE

Taking the~time-derivative of Eq. (\ref{tM}), regarding
Eqs. (\ref{tAnfngnd}), (\ref{tcu}), (\ref{tAj}),
and comparing with Eq. (\ref{Mu}), we arrive  at the~following
equation for function $B$:
\BE
\lvhz \frac{\tsn}{2B}\lvkz\frac{B,_\th}{\sn}\pvkz,_\th\pvhz,_\th=0\ .
\EE
The~general solution can be seen to be
\BE
B=a\sn\lvkz\frac{\sn}{\cs+1}\pvkz^{\rm C}
 +b\sn\lvkz\frac{\sn}{\cs+1}\pvkz^{-\rm C} \ ,\label{strB}
\EE
where $a$, $b$, ${\rm C}>0$ are constants. If we assume
${\rm C}\in[0,1]$ then $B$ is bounded %regular
at the~axis.

The~news function is obtained from (\ref{tcu}):
$c,_u=({\rm C}^2-1)/(2\dsn)$. As described in Ref. \cite{Bstruna}
and in Appendix~D  it corresponds
to the~news function of a string; we exclude $\rm C =0$.
%It is interesting to see explicitly
%It is described in detail in Appendix D that
Function $B$
given in Eq. (\ref{strB}) corresponds to translations along $z$-axis
and $t$-axis in the~spacetime of an infinite thin cosmic string
described by the~deficit angle $2\p (1-{\rm C})$, ${\rm C}\in(0,1]$; 
in the~weak-field
limit, ${\rm C}=1-4\m$, where $\m$ is the~mass per unit length
of the~string. (Notice that $\m\geq 0$ for ${\rm C}\in(0,1]$.)
%This is described in detail in Appendix D.
See Appendix~D for details.

In Appendix E, part 1, the~explicit form of electromagnetic field
for general $B$, Eq. (\ref{strB}), representing translations in
asymptotically flat spacetimes with a straight string is given.

If ${\rm C}=1$, there is no string extending to infinity.
Eq. (\ref{strB}) gives
\BE
B=(b-a)\cs+(b+a)\ .\label{trB}
\EE
Then $c,_u=0$ and from Eqs. (\ref{tAnfngnd}) and (\ref{tAn}) we get
\BE
\fn=(b-a)\sn\ ,\mm\An=\pul\lvhz -(b-a)\cs+b+a\pvhz\ .\label{trfnAn}
\EE
Consequently, regarding Eqs.
(\ref{Btetrada}), (\ref{Beta}) and (\ref{Afg}),
we find the~Killing vector field $\e^\a$ to be asymptotically of
the~form
\BE
\e^\a=[(b-a)\cs+b+a\ ,\ -(b-a)\cs+O(r^{-2})\ ,
\ (b-a)\sn\ \frac{1}{r}+O(r^{-3})\ ,\ O(r^{-3})]\ .\label{treta}
\EE
We thus see that the~Killing vector field
generates translations: with $b-a=0$
this is the~time translation, with $a+b=0$ the~translation along the~$z$-axis.

Since there is no string extending to infinity, both functions $c=c(\th)$
and $d=d(\th)$ are independent of time and both news functions
thus vanish; there is no radiation. By employing two
transformations from the~Bondi-Metzner-Sachs group, functions $c$
and $d$ can be transformed away.

For an illustration let us write down the~asymptotic form of both
gravitational and electromagnetic fields in the~case of
the~Killing vector representing timelike translations
in asymptotically flat spacetime without a string;
we thus assume $B=\mbox{const}$ and $c=d=0$. The~resulting
axially symmetric stationary metric and electromagnetic
field then have the~asymptotic form
\BEA
ds^2&=&\lvkz 1-\frac{2M(\th)}{r}+O(r^{-2})\pvkz du^2
          +2(1+O(r^{-4})) dudr
    +2O(r^{-1})dud\th+2O(r^{-1})\sn\ dud\f\nn\\
    &\ &-r^2\left[ (1+O(r^{-3}))d\th^2 +(1+O(r^{-3}))\dsn\ d\f^2
         +2O(r^{-3}) \sn\ d\th d\f\right]\ ,\\
\Fnj&=&-\frac{\ep_1}{r^2}
          +({e_1},_\th+e_1\ctg)\frac{1}{r^3}+O(r^{-4})\ ,\nn\\
\Fnd&=&O(r^{-2})\ ,\nn\\
\Fnt&=&O(r^{-2})\ ,\nn\\
\Fjd&=&\frac{e_1}{r^2}+O(r^{-3})\ ,\nn\\
\Fjt&=&f_1\sn\frac{1}{r^2}+O(r^{-3})\ ,\nn\\
\Fdt&=&-\m_1\sn-(f_1 \sn),_\th\frac{1}{r}+O(r^{-2})\ .\nn
\EEA
Using the~consequence of Einstein's equations (\ref{Nu}) and assuming
$N,_u=0$ (which, as will be proven in our forthcoming publication,
follows from further terms in $r^{-k}$
in the~Killing equations), we see that $M,_\th=0$, i.e.,
$\Aj(\th)=\mbox{const}$. Constants $M$, $\ep_1$ and $\m_1$
represent the~total mass, electric charge and magnetic charge,
respectively.

%% file: ajkap4.tex
\section{The~Boost Killing Vector}
In this section we shall find the~form of
the~gravitational news functions $c,_u$ and $d,_u$
for the~case of the~boost Killing vector by expanding the~Killing
equations in further orders in $r^{-1}$. We shall also obtain
the~form of the~electromagnetic news functions $X$ and $Y$.

We thus assume the~asymptotic form of the~Killing vector
to be given by Eq. (\ref{vboost}).
Expanding now the~Killing equations (\ref{Kill}) in higher orders
of $r^{-1}$ we obtain
\BEA
\ \L_\e \gnn &=&0\ \mm(r^{\ 0}):\ 2\An,_u+B,_u=
                      2\An,_u-\cs=0\ ,\label{BKjn}\\
\ \L_\e \gnj &=&0 \mm(r^{-1 }):\ \mbox{satisfied\ automatically}
                              \ ,\label{BKdmj}\\
\ \L_\e \gnd &=&0 \mm(r^{\ 1}):\ \fmj c,_u-\fn,_u+\Amj,_\th=
                               -\sn\ c,_u-\fn,_u-\sn=0
                                   \ , \label{BKtj}\\
\ \L_\e \gnt &=&0 \mm(r^{\ 1}):\ -\gen,_u+2\fmj d,_u=
                              -\gen,_u-2\sn\ d,_u=0\ ,\label{BKcj}\\
\ \L_\e \gjd &=&0 \mm(r^{\ 0}):\ \fn-\fmj c+B,_\th=
                               \fn+\sn\ c+u\sn=0\ ,\label{BKsn}\\
\ \L_\e \gjt &=&0 \mm(r^{\ 0}):\ 2\fmj d-\gen=
                            -2\sn\ d-\gen=0\ ,\label{BKsen}\\
\ \L_\e \gdd &=&0 \mm(r^{\ 1}):\ \fn,_\th+Bc,_u+\AnmB=
                  \fn,_\th-u\cs\ c,_u+\An+\pul u\cs =0
                    \ ,\label{BKoj}\\
\ \L_\e \gdt &=&0 \mm(r^{\ 1}):\ -\gen,_\th+\gen\ctg
                              -2Bd,_u-2d\fmj\ctg = \nn\\
              &\ & \mm\mv\mv    \  =
                   -\gen,_\th+\gen\ctg+2u\cs\ d,_u
                     +2d\cs=0 \ ,\label{BKdej}\\
\ \L_\e \gtt &=&0 \mm(r^{\ 1}):\ \pul B-\An+\fmj\cc-\fn\ctg+Bc,_u
                                         = \nn\\
              &\ & \mm\mv\mv    \  =
                       -\frac{u}{2}\cs-\An-\sn\cc
                -\fn\ctg-u\cs\ c,_u=0\ .\label{BKdsj}
\EEA
From Eqs. (\ref{BKjn}) and (\ref{BKsn}) we immediately get
\BEA
\An&=&\frac{u}{2}\cs+\r(\th)\ ,\label{49}\\
c  &=&-u-\frac{\fn}{\sn}\ ,\label{410}
\EEA
where  $\r(\th)$ is  an arbitrary function of $\th$.
Eq. (\ref{BKtj}) is just the~$u$-derivative of Eq. (\ref{BKsn}).
Eq. (\ref{BKoj}) implies
\BE
c,_u=\frac{1}{u\cs}(\fn,_\th+\An+\frac{u}{2}\cs)\ .\label{411}
\EE
Using now Eqs.  (\ref{49})-(\ref{411}) in  Eq.  (\ref{BKdsj})
we find that additive function $\r(\th)$ in Eq. (\ref{49})
must vanish, i.e.
\BE
\An=\frac{u}{2}\cs\ .\label{412}
\EE
Comparing Eqs. (\ref{BKtj}) and (\ref{BKoj}) we obtain an equation
for $\fn$,
\BE
u\fn,_u+\tg\ \fn,_\th+2u\sn=0\ .\label{rcefnb}
\EE
This equation can be solved by introducing a new variable
\BE
 w=\frac{\sn}{u}\ .\label{w}
\EE
Eq. (\ref{rcefnb}) then becomes
\BE
\der_u\fn (u,w)+2uw=0\ ,
\EE
so that the~general solution is
\BE
\fn=-u^2w+\K(w)=-u\sn+\K(\sn /u)\ ,\label{413}
\EE
where $\K(w)$ is an arbitrary function of $w$. Consequently,
Eq. (\ref{410}) leads to
\BE
c(u,\th)=-\frac{\K(w)}{uw}\ ,\label{414}
\EE
where $w$ is given by Eq. (\ref{w}).
The~news function thus reads
\BE
c,_u(u,\th)=\frac{\K(w),_w}{u^2}\ .\label{415}
\EE

Before writing down the~result for the~second news function,
let us compare the~expression (\ref{415}) with  the~news
function given for the~case of the~vacuum boost-rotation
symmetric spacetimes with the~hypersurface orthogonal
Killing vectors in Ref.~\cite{bicak}. There the~news function is
obtained in the~form $c,_u=F(\tilde U /\sn)/ \dsn$, where
$F$ is an arbitrary function of $\tilde U /\sn$  and
the~flat-space retarded time $\tilde U$ satisfies the~equation
$\tilde U,_u u\ctg=\tilde U\ctg-\tilde U,_\th$.
This equation can be rewritten as the~equation
$u\tilde U,_u+\tg\ \tilde U,_\th-\tilde U=0$, the~solution
of which is $\tilde U=\A (w)u$, $\A$ being function
of  $w=\sn/ u$. We can write
$c,_u=F(\A (w)/w)/ \dsn=
           F(\A (w)/w)/ (w^2u^2)=\K,_w(w)/ u^2$, where
$\K,_w=F(\A (w)/w)/ w^2$. Therefore, our result (\ref{415})
for the~general form of the~news function is in agreement
with Eq. (59)  given in Ref.~\cite{bicak}.

The~second news function, $d,_u$, can be found analogously.
Eq. (\ref{BKcj}) is the~$u$-derivative of Eq. (\ref{BKsen})
which gives
\BE
d=-\frac{\gen}{2\sn}\ .\label{416}
\EE
Substituting this result and $d,_u$ from Eq.
(\ref{BKcj})  into Eq. (\ref{BKdej}), we obtain
the~equation for $\gen$,
\BE
u\gen,_u+\tg\ \gen,_\th=0\ ,
\EE
which in terms of variables $u$ and $w$ simply yields
\BE
\gen=\gen(w)\ . \label{417}
\EE
The~second news function is thus given by
\BE
d,_u(u,\th)=\frac{\gen(w),_w}{2u^2}\ ,\label{dub}
\EE
where $\gen$ is an arbitrary function of $w=\sn /u$.

In order to obtain the~mass aspect and
the~total mass at null infinity we have to expand
the~Killing equations in higher orders in
$r^{-1}$. Straightforward though rather lengthy calculations
lead to the~following system of equations (in which
$\Amj$, $\An$, $B$, $\fmj$, $\fn$, $\gen$, $c$ and $d$ are
already known but are left unspecified
for the~sake of compactness):
\BEA
\ \L_\e \gnn &=&0 \mm(r^{-1 }):\ \Aj,_u-B,_uM+\Amj(cc,_u+dd,_u+M)
                               -\fmj M,_\th\nn\\
              &\ & \mv\mv\mv    -\dd(\gen,_u-2\fmj d,_u)
                                -\cc(\fn,_u-\fmj c,_u)=0
                                  \ ,\label{BKjmj}\\
\ \L_\e \gnj &=&0 \mm(r^{-2 }):\ \cc\ffc-\fmj(dd,_\th+cc,_\th)
                                         \nn\\
              &\ & \mv\mv\mv    -\frac{1}{2}B,_u(c^2+d^2)
                                   -\Aj-BM
                     -B(dd,_u+cc,_u)=0\ ,\label{BKdmd}\\
\ \L_\e \gnd &=&0 \mm(r^{\ 0}):\  \fmj(cc,_u+2dd,_u)
                                 -\fmj\cc,_\th-\fmj,_\th\cc\nn\\
              &\ & \mv\mv\mv  +\An,_\th+\pul B,_\th
                                 -\fn,_u c+\fn c,_u
                              -\fj,_u-2d\gen,_u=0\ ,\label{BKtn}\\
\ \L_\e \gnt &=&0 \mm(r^{\ 0}):\ -\gen c,_u+\gen,_uc-\gej,_u
                                    +2\fn d,_u+2\fmj(2dc,_u-cd,_u)\nn\\
              &\ & \mv\mv\mv   -\fmj\dd,_\th
                               -\fmj\dd\ctg=0 \ ,\label{BKcn}\\
\ \L_\e \gjd&=&0 \mm(r^{-1 }):\ 2\fj-2B\cc
                              -\fmj(c^2+2d^2)+2d\gen=0\ ,
                                   \label{BKsmj}\\
\ \L_\e \gjt&=&0 \mm(r^{-1 }):\ \gej-B\dd-d(\fn+\fmj c)=0 \ ,
                                      \label{BKsemj}\\
\ \L_\e \gdd
              &=&0 \mm(r^{\ 0}):\ 2d(\gen,_\th-\gen\ctg)
                                     +\Amj\frac{3d^2}{2}
                                +4\fmj d^2\ctg+\Aj
                                 +BM+\fj,_\th\nn\\
              &\ & \mv\mv\mv    +B(2dd,_u+cc,_u)-B\cc,_\th
                                    =0\ ,\label{BKon}\\
\ \L_\e \gdt
              &=&0 \mm(r^{\ 0}):\ -2\An d+Bd+4\fmj d(c,_\th+c\ctg)
                                -4\fn d\ctg\nn\\
              &\ & \mv\mv\mv   +\gen,_\th c-\gen(c,_\th+c\ctg)
                                 -\gej,_\th+\gej\ctg\nn\\
              &\ & \mv\mv\mv  +B\dd,_\th-B\dd\ctg=0\ ,\label{BKden}\\
\ \L_\e \gtt
              &=&0 \mm(r^{\ 0}):\ \Aj-\An c+\pul Bc+\pul\Amj(c^2+d^2)
                                   +BM-\fmj d^2\ctg\nn\\
              &\ & \mv\mv\mv     +2\fmj d\dd-B\ctg\cc
                                    +2B(cc,_u+dd,_u)\nn\\
              &\ & \mv\mv\mv    +\fj\ctg+\cc(-\fn+3\fmj c)
                            -c\ctg(\fn+\frac{3}{2}\fmj c)
                                     =0\ .\label{BKdsn}
\EEA
Substituting for $c$, $d$ and $\gen$ from Eqs.
(\ref{414}), (\ref{416}) and (\ref{417}) into Eqs.
(\ref{BKsmj}) and (\ref{BKsemj}) we obtain
\BE
\fj=\frac{1}{4\sn}(-2\K^2+{\gen}^2)+u\dctg (\K,_w w+\K)
          \ ,\label{427}
\EE
and
\BE
\gej=\frac{u}{2}\gen-\frac{\gen\K}{\sn}
           +\frac{u}{2}\dctg(\gen,_w w+\gen)\ .\label{428}
\EE
Using the~results for $\fmj$, $\fn$, $B$, ... already obtained,
we find that Eqs. (\ref{BKtn}), (\ref{BKcn}) and (\ref{BKden})
are just identities;
Eqs. (\ref{BKon}), (\ref{BKdsn}) become also identities if
Eq. (\ref{BKdmd}) is used.

Comparing the~remaining Eq. (\ref{BKjmj}) with the~$u$-derivative of
Eq.~(\ref{BKdmd}), we finally arrive at the~equation
for the~mass aspect $M$:
\BE
uM,_u+\tg M,_\th +3M-u\tg\cc,_u=0\ ,
\EE
where $c,_u$ is given by Eq. (\ref{415}). This equation
can be solved by the~substitution (\ref{w}) to yield
\BE
M(u,\th)=\frac{1}{2u}(\K,_{ww}w+2\K,_w)+\frac{\L(w)}{u^3}\ ,\label{bM}
\EE
where $\L(w)$ is an arbitrary function of $w$.
Substituting $M$ into
Eq. (\ref{BKdmd}) we obtain  function $\Aj$ in the~form
\BE
\Aj=\cs \lvkz\pul\K,_{ww}w+2\K,_w+\frac{\K}{w}\pvkz
      -\frac{\cs}{8\dsn}\lvkz 4\K^2+{\gen}^2\pvkz+\frac{\cs}{u^2}\L(w)
         \ .\label{bAj}
\EE
The~last expression for $M$  can be written as
\BE
M(u,\th)=\frac{1}{2\sn}(w^2\K,_w),_w+\frac{\L}{u^3}\ .\label{bbM}
\EE
Comparing this form of $M(u,\th)$ with the~consequence
of Einstein's equations  (\ref{Mu}), we find
\BE
\L(w)=\frac{\l(w)}{w^3}\ ,
\EE
where $\l$ has to satisfy the~equation
\BE
\l(w),_w=w^2(\K,_w^2+\frac{1}{4}\gen,_w^2+\E^2+\tilde{\BB}^2)
-\frac{1}{2w}(w^3\K,_{ww}),_w\ .\label{boostlambda}
\EE
Here, $\K$ and $\gen$ determine the~gravitational news functions,
$c,_u$ and $d,_u$, by relations (\ref{415}) and (\ref{dub}),
$\E$ and $\tilde{\BB}$ determine the~electromagnetic news functions,
$X$ and $Y$, given below by relations (\ref{E24}) and (\ref{E25}).
Hence, solving the~last equation for $\l$ for  {\it{given}}
$\K$, $\gen$, $\E$ and $\tilde{\BB}$, we find $\L(w)$ and thus
the~mass aspect $M(u,\th)$ in the~form of Eq. (\ref{bbM}).
The~total mass at
$\I^+$ is then given by integrating Eq. (\ref{bbM}) over the~sphere:
\BE
m(u)=\pul\int_{0}^{\p} M(u,\th)\sn d\th\
     =\frac{1}{4}\int_0^\p (w^2\K,_w),_wd\th
     +\pul\int_0^\p \frac{w\L}{u^2}d\th\ .
\EE

Substituting the~expansions of the~metric functions, 
Eq. (\ref{rozvmetr}),
into the~null tetrad, Eq. (\ref{Btetrada}), and
coefficients $A$, $B$, $\tilde{f}$ and $\tilde{g}$, Eqs. (\ref{412}), (\ref{413}),
(\ref{414}), (\ref{416}), (\ref{417}), (\ref{427}), (\ref{428}),
(\ref{bM}), (\ref{bAj}), into Eq.~(\ref{Beta}), we find the~expansion
of the~boost Killing vector to be
\BEA
\e^\m&= &
      \lvhz -u\cs\ ,\ r\cs+u\cs+\cs\ \lvkz\K,_w +\frac{\K}{w}\pvkz
                      \ \frac{1}{r}
              +O(r^{-2})\ ,\ \label{etaboost}\\
     &\ &\mm
             -\sn-u\sn\ \frac{1}{r} + uc\sn\ \frac{1}{r^2}+O(r^{-3})\ ,
               \ ud\ \frac{1}{r^2}+O(r^{-3})\pvhz\ .\nn
\EEA
It can easily be seen that the~boost Killing vector generating
Lorentz transformations along the~$z$-axis in Minkowski space,
$\e^\a_M =[z\ ,\ 0\ ,\ t\ ,\ 0]$, in \{ $u$,~$r$,~$\th$,~$\f$ \}
coordinates reads
$\e^\a_M =[-u\cs\ ,\ r(1+u/r)\cs\ ,\ -\sn(1+u/r)\ ,\ 0]$
so that both vectors differ only
in  terms proportional to $\K$, $c$ and $d$.

Let us finally turn  to the~asymptotic properties
of the~electromagnetic field which now is assumed to satisfy
$\L_\e \Fab=0$, Eq. (\ref{E1}), where $\e$ is the~boost Killing field
(\ref{etaboost}). The~first orders of these equations read
\BEA
\L_\e \Fnj&=&0\mm (r^{- 2}):\mm
                    \cs(\tg\ep,_\th+u\ep,_u+2\ep+u\tg X)=0
                    \ ,\label{E18}\\
\L_\e \Fnd&=&0\mm (r^{\ 0}):\mm -\cs(\tg X,_\th+uX,_u+2X)=0
                    \ ,\label{E19}\\
\L_\e \Fnt&=&0\mm (r^{\ 0}):\mm
                     -\cs\lhz\tg(Y\sn),_\th+u(Y\sn),_u+Y\sn\phz=0
            \ ,\label{E20}\\
\L_\e \Fjd&=&0\mm (r^{- 2}):\mm
                -\cs(\tg e,_\th+ue,_u+2e-u\tg\ep)=0\ ,\label{E21}\\
\L_\e \Fjt&=&0\mm (r^{- 2}):\mm
          -\cs\lhz\tg(f\sn),_\th+u(f\sn),_u+f\sn
                  +u\tg(\m\sn)\phz=0\ ,\label{E22}\\
\L_\e \Fdt&=&0\mm (r^{\ 0}):\mm
          \cs\lhz\tg(\m\sn),_\th+u(\m\sn),_u+\m\sn
                      +u\tg(Y\sn)\phz =0\ .\label{E23}
\EEA
Using again variable $w$ given by Eq. (\ref{w}),
Eqs. (\ref{E19}) and (\ref{E20})
can easily be solved to yield
\BEA
X(u,\th)&=&\frac{\E(w)}{u^2}\ ,\label{E24}\\
Y(u,\th)&=&\frac{\BB(w)}{u\sn}=\frac{\BB(w)}{wu^2}=\frac{\tilde{\BB}(w)}{u^2}
    \ ,\label{E25}
\EEA
where functions $\E(w)$ and $\BB(w)$ are arbitrary
integration functions
and $\tilde{\BB}(w)=\BB(w)/w$. Substituting these results into
Eqs. (\ref{E18}) and (\ref{E23}) we find
\BEA
\ep&=&\frac{\E(w)       }{  u}\ctg+\frac{\F(w)       }{u^2}
                        \ ,\label{E26}\\
\m &=&\frac{\BB(w)       }{\sn}\ctg+\frac{\C(w)       }{u\sn}
    = \frac{\tilde{\BB}(w)}{  u}\ctg+\frac{\tilde{\C}(w)}{u^2}
                         \ ,\label{E27}
\EEA
where functions $\F(w)$ and $\C(w)$ are arbitrary integration functions
and $\tilde{\C}(w)=\C(w) /w$. Then the~solution of
Eqs. (\ref{E21}) and (\ref{E22}) reads
\BEA
e&=&\frac{\E(w)}{2}-\frac{\F(w)}{u}\ctg+\frac{\GG (w)}{u^2}
        \ ,  \label{E28}\\
f&=&-\frac{\BB (w)u}{2\sn}+\frac{\C(w)}{\sn}\ctg+\frac{\D (w)}{u\sn}
  = -\frac{\tilde{\BB}(w)}{2}+\frac{\tilde{\C}(w)}{u}\ctg
           +\frac{\tilde{\D}(w)}{u^2}\ , \label{E29}
\EEA
where $\GG(w)$ and $\D(w)$ are again integration functions and
$\tilde{\D}(w)=\D (w)/w$.
Comparing Eqs. (\ref{E18}) and (\ref{E23})
with Maxwell equations (\ref{uderep}) and (\ref{udermi}), and
using previous results,
we get equations for $\F(w)$ and $\C(w)$:
$\F,_w w+2\F=0$ and $\C,_w w+\C=0$. Their solutions are
\BEA
\F&=&\frac{l}{w^2}\ ,\mm l=\mbox{const}\ ,\label{E30}\\
\C&=&\frac{p}{w}\ ,\mm p=\mbox{const}\ .\label{E31}
\EEA
Since $w=\sn /u$, the~regularity of the~electromagnetic field
implies $l=p=0$.

Finally, comparing  Maxwell equations (\ref{udere}) and  (\ref{uderf})
with Eqs. (\ref{E21}) and (\ref{E22}), we get restrictions
on functions $\D(w)$ and $\GG(w)$ in the~form
\BEA
\E,_w w-\E+2w^2(\GG,_w w+2\GG)+2w\E\K+\BB\gen &=& 0\ ,\\
-(\BB,_w w-2\BB)+2w^2(\D,_w w+\D)-2w\BB\K+w^2\gen\E &=& 0\ .
\EEA

%% file: ajap1.tex
In our convention (see the~end of Introduction)
the~Einstein-Maxwell equations read
\BE
K_{\m\n}  \rov\Rmn+8\p\Tmn=0\ ,\label{Einrce}
\EE
where the~electromagnetic stress tensor is given by
\BE
\Tmn=\frac{1}{4\p}(F_{\m\s}F_\n^{\ \s}
       -\frac{1}{4}\gmn F_{\s\r}F^{\s\r})\ ,
\EE
and $T^\m_\m=R=0$; the~Maxwell equations are
\BEA
G_{\m\n\l}&\rov&F_{[\m\n,\l]\text{cykl.}}=0\ ,\label{G}\\
J^\m      &\rov&\Fumn_{\mm;\n}=0\ .  \label{J}
\EEA
In paper \cite{burg} the~tensor $\Emn =4\p\Tmn$ is introduced;
however, the~field equations (\ref{Einrce}) are erroneously
written without the~factor $2$ at $\Emn$.
This can be "cured" by considering
$F_{\m\n\text{(Burg)}}=\sqrt{2} F_{\m\n\text{(real)}}$.
Following \cite{burg}, eighteen equations (\ref{Einrce}),
(\ref{G}) and (\ref{J}) can be
divided into twelve main equations, one trivial equation, and
five supplementary conditions. The~main equations are
\BEA
&\ &\Kjj  =\Kjd=\Kjt=\Kdd=\Kdt=\Ktt=0\ ,\\  %\label{BhlrK}\\
&\ &\GGjdt=\GGnjd=\GGnjt=0\ ,\ \Jn=\Jd=\Jt=0\ . \nn  %\label{BhlrGJ}
\EEA
If these are satisfied, then also
\BE  \Knj=0 \ .\EE
The~only further equations to be satisfied are
the~supplementary conditions
\BEA
(r^2\Knn),_r&=&(r^2\Knd),_r=(r^2\Knt),_r=0
        \label{Budodpod}\ ,\\  %\label{BdodpK}\\
\GGndt,_r   &=&0\ ,\mm\Jj,_r=0\ ,\nn      %\label{BdodpGJ}
\EEA
which imply that $r^2(\Knn, \Knd, \Knt )$, $\GGndt$, and
$\Jj$ are functions of $u$, $\th$, $\f$ only.

Since in \cite{burg} all Einstein equations contain errors
due to the~factor $2$ mentioned above, and equations
(5), (6), (7), (14), (15), (17), (18), (19) in \cite{burg}
contain additional errors, we write all field equations explicitly
here (we checked them by using MAPLE V).
Starting from the~metric of the~form (\ref{ds}) we find that
the~main field equations are (denoting $ch=\chd$, $sh=\shd$)
\BEA
&\  &\frac{r}{4}\Kjj=0\ :\\
&\  &\mm\br=\frac{r}{2}\lvkz\gr^2 ch^2+\dr^2\pvkz
      +\frac{1}{2r}\lvkz\emdg ch\Fjd^2+\edg ch\Fjt^2\dcsec
            -2sh\Fjd\Fjt\csec\pvkz\ ,\nn\\
&\  &2r^2\Kjd=0\ :\\
&\  &\mm  \lvsz r^4\emdb(\edg\Ur ch +\Wr sh)\pvsz,_r=\nn\\
&\  &\mv\mv\mm =2r^2\lvsz\brt-\frac{2}{r}\bth+2\dr\dth
                -4\gr\dth sh ch
               -(\grt+2\gr\ctg-2\gr\gth)ch^2\pvsz\nn\\
&\  &\mv\mv\mm +2r^2\edg\csec\lvsz  -\drf-2\dr\gf
             +2\gr\df(1+2sh^2)+(\grf+2\gr\gf)sh ch\pvsz\nn\\
&\  &\mv\mv\mm -4r^2\emdb\FUW\Fjd
             -4(sh \Fjd-\edg ch\Fjt\csec)\Fdt\csec\ ,
               \nn\\
&\  &\frac{2r^2}{\sn}\Kjt=0\ :\\
&\  &\mm\lvsz r^4\emdb(\emdg\Wr ch+\Ur sh)\pvsz,_r =
            2r^2\emdg\lvsz -\drt+2\dr\gth-2\dr\ctg\nn\\
&\  &\mv\mv\mm  -(\grt+2\gr\ctg-2\gr\gth)sh ch
            -2\gr\dth (1+2 sh^2)\pvsz\nn\\
&\  &\mv\mv\mm +2r^2\csec\lvsz \brf -\frac{2}{r}\bfi+2\dr\df
            +4\gr\df sh ch+(\grf +2\gr\gf)ch^2\pvsz\nn\\
&\  &\mv\mv\mm -4r^2\emdb\FUW\Fjt\csec
          -4(\emdg ch\Fjd-sh\Fjt\csec)\Fdt\csec\ ,\nn\\
&\  &\frac{1}{2}\edb\lvsz (\emdg\Kdd +\edg\dcsec\Ktt)ch
                 -2\csec \Kdt sh\pvsz=0\ :\\
&\  &\mm   \Vr =2\edb\csec\lvsz (\btf +\bth\bfi+2\dth\df)sh
                     +(\dtf+\df\ctg+\dth\gf-\gth\df
                      +\dth\bfi+\bth\df)ch\pvsz  \nn\\
&\  &\mv\mv\mm -\edbmg\lvsz(\bthth+\bth^2+\bth\ctg
              +2\gth^2+2\dth^2%\nn\\
%&\  &\mv\mv\mm
               -1-\gthth-3\gth\ctg-2\bth\gth)ch\nn\\
&\  &\mv\mv\mm +(\dthth+3\dth\ctg+2\bth\dth-4\gth\dth)sh\pvsz\nn\\
&\  &\mv\mv\mm -\edbg\lvsz(\bff+\bfi^2+2\gf^2+2\df^2
              +\gff+2\bfi\gf)ch
                +(\dff+2\bfi\df+4\gf\df)sh\pvsz\dcsec\nn\\
&\  &\mv\mv\mm  -\frac{r^4}{4}
                  \emdb\lvsz(\edg\Ur^2+\emdg\Wr^2)ch
                     +2\Ur\Wr sh\pvsz \nn\\
&\  &\mv\mv\mm +\frac{r}{2}(r\Urt+r\Ur\ctg+4\Uth+4U\ctg)
                 +\frac{r}{2}\csec(r\Wrf+4\Wf)\nn\\
&\  &\mv\mv\mm -\frac{1}{r^2}\edb
           \lvsz [r^2\emdb\FUW ]^2+(\Fdt\csec)^2\pvsz\ ,
                 \nn\\
&\  &\frac{1}{4r}\edb(\emdg\Kdd-\edg\Ktt\dcsec)=0\ :\\
&\  &\mm (r\g),_{ur}ch +2r(\gu\dr+\du\gr)sh
           =\pul(\gr\Vr+\grr V+\frac{1}{r}\gr V)ch
               + 2\gr\dr V sh\nn\\
&\  &\mv\mv\mm +\frac{r^3}{8}\emdb (\edg\Ur^2-\emdg\Wr^2)
              +\frac{1}{2r}\edbmg(\bthth+\bth^2-\bth\ctg)\nn\\
&\  &\mv\mv\mm -\frac{1}{2r}\edbg(\bff+\bfi^2)\dcsec
                +\frac{1}{r}\edb(\bth\df-\bfi\dth)\csec\nn\\
&\  &\mv\mv\mm  +\frac{r}{4}\edg\csec
                  \lvsz (\Urf+\frac{2}{r}\Ufi )sh
                  +4\dr\Ufi ch\pvsz\nn\\
&\  &\mv\mv\mm -\frac{r}{4}\emdg\lvsz
                (\Wrt-\Wr\ctg+\frac{2}{r}\Wth
                -\frac{2}{r}W\ctg)sh
              +4\dr(\Wth-W\ctg)ch\pvsz\nn\\
&\  &\mv\mv\mm -\frac{1}{4}(r\Urt+2\Uth-r\Ur\ctg-2U\ctg+4\gth U
                  \nn\\
&\  &\mv\mv\mm +4r\grt U+2r\gth\Ur
               +2r\gr\Uth+2r\gr U\ctg)ch\nn\\
&\  &\mv\mv\mm -r(\dr\Uth+2\gr\dth U+2\dr\gth U
                -\dr U\ctg)sh\nn\\
&\  &\mv\mv\mm +\frac{1}{4}\csec(r\Wrf+2\Wf-4\gf W
               -4r\grf W-2r\gf\Wr-2r\gr\Wf)ch\nn\\
&\  &\mv\mv\mm +r\csec(\dr\Wf-2\dr\gf W-2\gr\df W)sh\nn\\
&\  &\mv\mv\mm -\frac{1}{2r}\lvsz 2
                  (\emdg\Fnd\Fjd-\edg\Fnt\Fjt\dcsec)
                  -\frac{1}{r}V (\emdg\Fjd^2-\edg\Fjt^2\dcsec)\nn\\
&\  &\mv\mv\mm  -2(\emdg W\Fjd+\edg U\Fjt\csec)\Fdt\csec
                 \pvsz\ ,\nn\\
&\  &\frac{1}{4r}\edb\lvsz
           (\emdg\Kdd+\edg\csec\Ktt)sh
          -2\csec\Kdt ch\pvsz =0\ :\\
&\  &\mm (r\d),_{ur}-2r\gu\gr sh ch =
              \pul(\dr\Vr+\drr V
              +\frac{1}{r}\dr V-2\gr^2 V sh ch)\nn\\
&\  &\mv\mv\mm +\frac{r^3}{8}\emdb\lvsz
                (\edg\Ur^2+\emdg\Wr^2)sh
                   +2\Ur\Wr ch\pvsz\nn\\
&\  &\mv\mv\mm -\frac{1}{2r}\edbmg
                 (\bthth+\bth^2-\bth\ctg) sh
                 -\frac{1}{2r}\edbg
                  (\bff+\bfi^2) sh\dcsec\nn\\
&\  &\mv\mv\mm -\frac{1}{r}\edb
                (-\btf-\bth\bfi+\bfi\ctg
                +\bth\gf-\bfi\gth) ch\csec\nn\\
&\  &\mv\mv\mm -\frac{r}{2}\lvsz
                   2\drt U+\frac{2}{r}\dth U
                  +\dr\Uth+\dth\Ur+\dr U\ctg
                -2\gr(\Uth-U\ctg+2\gth U)sh ch\pvsz\nn\\
&\  &\mv\mv\mm -\frac{r}{2}\csec\lvsz
                2\drf W+\frac{2}{r}\df W+\dr\Wf
                +\df\Wr+2\gr(\Wf-2\gf W)sh ch\pvsz\nn\\
&\  &\mv\mv\mm -\frac{r}{4}\emdg\lvsz
                 \Wrt-\Wr\ctg+\frac{2}{r}\Wth
                  -\frac{2}{r}W\ctg-4\gr(\Wth-W\ctg)ch^2\pvsz\nn\\
&\  &\mv\mv\mm
                -\frac{r}{4}\edg\csec
                 (\Urf+\frac{2}{r}\Ufi+4\gr\Ufi ch^2)\nn\\
&\  &\mv\mv\mm -\frac{1}{2r}\lvsz
               2(ch\Fnd-\edg sh\Fnt\csec)\Fjt\csec
                  -2(\emdg sh\Fnd-ch\Fnt\csec)\Fjd\nn\\
&\  &\mv\mv\mm +\frac{1}{r}V(\emdg sh\Fjd^2+
             \edg sh\Fjt^2\dcsec-2ch\Fjd\Fjt\csec)\nn\\
&\  &\mv\mv\mm +2[U(ch\Fjd-\edg sh\Fjt\csec)
                +W(\emdg sh\Fjd-ch\Fjt\csec)]\Fdt\csec\pvsz\ ,\nn\\
&\  &\GGjdt=0\ :\\
&\  &\mm       \Fdt,_r=\Fjt,_\th-\Fjd,_\f\ ,\nn\\
&\  &\GGnjd=0\ :\\
&\  &\mm       \Fnj,_\th=\Fnd,_r-\Fjd,_u\ ,\nn\\
&\  &\GGnjt=0\ :\\
&\  &\mm       \Fnj,_\f=\Fnt,_r-\Fjt,_u\ ,\nn\\
&\  &r^2\edb\Jn=0\ :\\
&\  &\mm       \lvsz r^2\emdb\FUW\pvsz,_r=\nn\\
&\  &\mv\mv\mm  =-\csec\lvsz\sn
             (\emdg ch\Fjd-sh\Fjt\csec)\pvsz,_\th
           +\csec(sh\Fjd-\edg ch\Fjt\csec),_\f\ ,\nn\\
&\  &r^2\edb\Jd=0\ :\\
&\  &\mm      (\emdg ch\Fjd-sh\Fjt\csec),_u
              +(\emdg ch\Fnd-sh\Fnt\csec),_r=\nn\\
&\  &\mv\mv\mm=\lvsz r^2\emdb U\FUW\nn\\
&\  &\mv\mv\mm+\V(\emdg ch\Fjd-sh\Fjt\csec)
               +(\emdg ch W+sh U)\Fdt\csec\pvsz,_r\nn\\
&\  &\mv\mv\mm -\csec\lvsz W(\emdg ch\Fjd -sh\Fjt\csec)
               +U(sh\Fjd-\edg ch\Fjt\csec)
           +\frac{1}{r^2}\edb\Fdt\csec\pvsz,_\f\ ,\nn\\
&\  &r^2\sn\edb\Jt=0\ :\\
&\  &\mm       (sh\Fjd-\edg ch\Fjt\csec),_u
              +(sh\Fnd-\edg ch\Fnt\csec),_r=\nn\\
&\  &\mv\mv\mm =-\lvsz r^2\emdb W\FUW\nn\\
&\  &\mv\mv\mm -\V(sh\Fjd-\edg ch\Fjt\csec)
                -(\edg ch U+sh W)\Fdt\csec\pvsz,_r\nn\\
&\  &\mv\mv\mm  -\lvsz W(\emdg ch\Fjd-sh\Fjt\csec)
                  +U(sh\Fjd-\edg ch\Fjt\csec)
             +\frac{1}{r^2}\edb\Fdt\csec\pvsz,_\th\ .\nn
\EEA
Since we assume the~axial symmetry, hereafter we put all derivatives
$\der /\der\f$ equal to zero.

Now following \cite{burg}, we assume functions $\g$, $\d$, $\Fjd$
and $\Fjt$ to have the~expansions at large $r$ on a hypersurface
$u=u_0$ of the~form
\BEA
\g  &=&\frac{c}{r}+\lvkz C-\frac{1}{6}c^3-\frac{3}{2}cd^2\pvkz
           \frac{1}{r^3}+\frac{D}{r^4}
         +O(r^{-5})\ ,\\  %\label{gama}\\
\d  &=&\frac{d}{r}+
           \lvkz H-\frac{1}{6}d^3+\frac{1}{2}c^2 d\pvkz\frac{1}{r^3}
                  +\frac{K}{r^4}+O(r^{-5})\ ,\label{HRN}\\
\Fjd&=&\frac{e}{r^2}+\lvkz 2E+ec+fd\pvkz \frac{1}{r^3}
             +O(r^{-4})\ ,\\  %\label{Fjd}
\Fjt&=&\lvvkz\frac{f}{r^2}+\lvkz 2F+ed-fc\pvkz \frac{1}{r^3}
              +O(r^{-4})\pvvkz\sn\ ,
\EEA
where  $c$, $C$, $d$, ... are prescribed functions
of $\th$ on $u=u_0$. This corresponds to the~outgoing
radiation condition and the~absence of logarithmic terms.
(We thus do not consider polyhomogeneous null infinity of
Ref.~\cite{polyhom}.)
Then twelve main equations determine the~other eight
functions $\b$, $U$, $V$, $W$, $\Fnj$, $\Fnd$, $\Fnt$
and $u$-derivatives
of the~four prescribed functions
$\g$, $\d$, $\Fjd$ and $\Fjt$ on the~hypersurface $u=u_0$.
Denoting by $N$, $P$, $M$, $\ep$, $\m$, $X$ and $Y$
arbitrary functions of $\th$ on $u=u_0$, we find
\BEA
\b  &= & -\frac{1}{4}(c^2+d^2)\frac{1}{r^2}+O(r^{-4})\ ,
                \\  %\label{beta}\\
U   &= &-\cc\frac{1}{r^2}
        +\lvkz 2N+3(cc,_\th+dd,_\th)+4(c^2+d^2)\ctg\pvkz
           \frac{1}{r^3}\nn\\  %\label{U}\\
    &\ &+\pul\lvkz 3(C,_\th+2C\ctg)-6(cN+dP)
            -4(2c^2c,_\th+cdd,_\th+c,_\th d^2)\pvkz\frac{1}{r^4}\\
    &\ &+\pul\lvkz -8c(c^2+d^2)\ctg
          +2(\ep e-f\m)\pvkz\frac{1}{r^4}+O(r^{-5})\ ,\nn\\
W   &= &-\dd\frac{1}{r^2}
        +\lvkz 2P+2(c,_\th d-cd,_\th)\pvkz\frac{1}{r^3}\nn\\
    &\ &+\pul\lvkz 3(H,_\th+2H\ctg)+(cP-dN)
                 -4(2d^2d,_\th+cdc,_\th+c^2d,_\th)\pvkz\frac{1}{r^4}\\
    &\ &+\pul\lvkz -8d(c^2+d^2)\ctg+2(\m e+\ep f)\pvkz\frac{1}{r^4}
                          +O(r^{-5})\ ,\nn\\
V   &= &r-2M
        -\lvkz N,_\th+N\ctg-\pul (c^2+d^2)\pvkz\frac{1}{r}\nn\\
    &\ &-\lvkz -\cc^2-\dd^2 -(\ep^2+\m^2)\pvkz\frac{1}{r}\nn\\
    &\ &-\pul
         \lvkz C,_{\th\th}+3C,_\th\ctg-2C+6N\cc
                                 +6P\dd\pvkz\frac{1}{r^2}\\
    &\ &-\pul
         \lvkz 4(2cc,_\th^2+3c,_\th dd,_\th-cd,_\th^2)
           +8(2c,_\th d^2+3c^2c,_\th+cdd,_\th)\ctg\pvkz
                  \frac{1}{r^2}\nn\\
%\label{HRF}\\
    &\ &-\pul
         \lvkz 16c(c^2+d^2)\dctg
             +2\ep\ee-2\m (f,_\th+f\ctg)\pvkz\frac{1}{r^2}
                +O(r^{-3})\ ,\nn\\    %\label{V}\\
\Fnj&=&-\frac{\ep}{r^2}+\lvkz e,_\th+e\ctg\pvkz\frac{1}{r^3}
          +O(r^{-4})\ ,\\   %\label{Fnj}\\
\Fdt&=&\lvvkz -\m-\lvkz f,_\th+f\ctg\pvkz\frac{1}{r}+O(r^{-2})\pvvkz\sn\ ,\\
\Fnd&=&X+\lvkz \ep,_\th-e,_u\pvkz\frac{1}{r}
        -\lvkz [E+\pul (ec+fd)],_u
              +\pul\ee,_\th\pvkz\frac{1}{r^2}
                  +O(r^{-3})\ ,\\    %\label{Fnd}
\Fnt&=&\lvvkz Y-\frac{f,_u}{r}
        -\lvkz [F+\pul(ed-fc)],_u\pvkz\frac{1}{r^2}
                   +O(r^{-3})\pvvkz\sn\ ,
\EEA
and
\BEA
2e,_u &=&\ep,_\th-2(cX+dY)\ ,\label{Bueu}\\
2f,_u &=&-\m,_\th-2(-cY+dX)\ ,\\
4E,_u &=&-\lvkz\frac{\der}{\der\th}+2\ctg\pvkz
        \lvhz\lvkz\frac{\der}{\der\th}-\ctg\pvkz e
                +2(c\ep+d\m)\pvhz\ ,\\
4F,_u &=&-\lvkz\frac{\der}{\der\th}+2\ctg\pvkz
        \lvhz\lvkz\frac{\der}{\der\th}-\ctg\pvkz f
                +2(d\ep-c\m)\pvhz\ ,\\
4C,_u &=&2(c^2-d^2)c,_u+4dcd,_u+2cM
           +d\lvkz\frac{\der}{\der\th}+\ctg\pvkz\dd\nn\\
      & &\ -\lvkz\frac{\der}{\der\th}-\ctg\pvkz N+2(eX-fY)
           \ ,\\  %\label{HRU}\\
4H,_u &=&-2(c^2-d^2)d,_u+4dcc,_u+2dM
           -c\lvkz\frac{\der}{\der\th}+\ctg\pvkz\dd\nn\\
      & &\  -\lvkz\frac{\der}{\der\th}-\ctg\pvkz P+2(eY+fX)\ ,\\
4D,_u &=&\lvkz\frac{\der}{\der\th}+3\ctg\pvkz
         \lvhz-\lvkz\frac{\der}{\der\th}-2\ctg\pvkz C
                    +2(cN-dP)\pvhz\nn\\
      & &\ -\frac{2}{3}\ep\lvkz\frac{\der}{\der\th}-\ctg\pvkz e
           +\frac{2}{3}\m \lvkz\frac{\der}{\der\th}-\ctg\pvkz f\\
      & &\ -\frac{4}{3}c(\ep^2+\m^2)+\frac{8}{3}(EX-YF)\ ,\nn\\
4K,_u &=&\lvkz\frac{\der}{\der\th}+3\ctg\pvkz
        \lvhz-\lvkz\frac{\der}{\der\th}-2\ctg\pvkz H
                   +2(cP+dN)\pvhz\nn\\
     & &\ -\frac{2}{3}\ep\lvkz\frac{\der}{\der\th}-\ctg\pvkz f
          -\frac{2}{3}\m \lvkz\frac{\der}{\der\th}-\ctg\pvkz e
         \label{BuKu}\\
     & &\ -\frac{4}{3}d(\ep^2+\m^2)+\frac{8}{3}(FX+YE)\ .\nn
\EEA
In order to simplify the~last equations one can, following
\cite{burg},  introduce ten new quantities
\BE
\begin{array}{ccc}
c^*=c+id\ ,\ \mv\mv\mv\mv\mv\mm&  X^*  =X  +iY \ ,&\ \\[1mm]
C^*=C+iH\ ,  \mv\mv\mv\mv\mv\mm&  e^*  =e  +if \ ,&\ \\[1mm]
D^*=D+iK\ ,  \mv\mv\mv\mv\mv\mm&  E^*  =E  +iF \ ,&\ \\[1mm]
N^*=N+iP\ ,  \mv\mv\mv\mv\mv\mm&  \ep^*=\ep+i\m\ ,&\ \\[1mm]
M^*=M
  +i\pul\lvkz\frac{\der}{\der\th}
  +\ctg\pvkz\dd\ , &   \  &\ \\[1mm]
{\cal L}_p=-\lvkz\frac{\der}{\der\th}-p\ctg\pvkz\ ,\ p=-3, -2...\ 2
                    \ ,&   \  &\ \\
\end{array}  \\
\label{komplexnic}
\EE
in terms of which equations (\ref{Bueu})-(\ref{BuKu})
become
\BEA
2e^*,_u  &=&-{\cal L}_0\bar\ep^* -2c^* \bar X^*\ ,\\
4E^*,_u  &=&-{\cal L}_{-2}({\cal L}_1 e^*-2c^*\bar\ep^*)\ ,
                \label{komplrce}\\
4C^*,_u  &=&2c^{*2}\bar c^*,_u+2c^*\bar M^*
           +{\cal L}_1 N^*+2e^*X^*\ ,\\
4D^*,_u  &=&-{\cal L}_{-3}({\cal L}_2 C^*+2c^*N^*)
           +\frac{2}{3}\ep^*{\cal L}_1e^*
        -\frac{4}{3}c^*\ep^*\bar\ep^*+\frac{8}{3}E^*X^*\ .
\EEA
The~supplementary conditions (\ref{Budodpod}) have the~form
\BEA
M,_u  &=&-(c,_u^2+d,_u^2)-(X^2+Y^2)
         +\pul(c,_{\th\th}+3c,_\th\ctg-2c),_u\ ,\\
%     & &\ +\frac{1}{2}\lvkz\frac{\der}{\der\th}+\ctg\pvkz
%           \lvkz\frac{\der}{\der\th}+2\ctg\pvkz c_0\\
3N,_u &=&-M,_\th-2c\lvkz\frac{\der}{\der\th}+2\ctg\pvkz c,_u
             -2d\lvkz\frac{\der}{\der\th}+2\ctg\pvkz d,_u\label{Nu}\\
      & &\ -(cc,_u),_\th-(dd,_u),_\th
           -2(\ep X+\m Y)\ ,\nn\\   %\label{dodat}\\
3P,_u &=&\pul\frac{\der}{\der\th}
          \lvkz\frac{\der}{\der\th}+\ctg\pvkz\dd
           +2c\lvkz\frac{\der}{\der\th}+2\ctg\pvkz d,_u
           -2d\lvkz\frac{\der}{\der\th}+2\ctg\pvkz c,_u\\
      & &\ +(cd,_u),_\th-(dc,_u),_\th -2(\ep Y-\m X)\ ,\nn\\
\ep,_u&=&-X,_\th-X\ctg\ ,\\
%         \lvkz\frac{\der}{\der\th}+\ctg\pvkz X\\
\m,_u &=&-Y,_\th-Y\ctg\ ,
\EEA
which in terms of quantities (\ref{komplexnic}) simplify to
\BEA
M^*,_u  &=&-c^*,_u\bar c^*,_u-X^*\bar X^*
           +\pul\L_{-1}\L_{-2}c^*,_u\ ,\\
3N^*,_u &=& \L_0\bar M^*+2c^*\L_{-2}\bar c^*,_u
          +\L_0(c^*\bar c^*,_u)
          -2\bar\ep^* X^*\ ,\label{kompldodat}\\
\ep^*,_u&=&{\cal L}_{-1}X^*\ .
\EEA

The~structure of the~field equations is thus  following.
Nine functions $\g$, $\d$, $\Fjd$, $\Fjt$,
$M$, $N$, $P$, $\ep$ and $\m$ are  prescribed on
an "initial" hypersurface
$u=u_0$, and four functions $c,_u$, $d,_u$, $X$ and $Y$
have to be prescribed
for all $u$. Then the~time evolution of
gravitational and electromagnetic  fields is fully determined.
Functions $c,_u$ and $d,_u$ are the~well-known {\it gravitational
news  functions}, functions $X$ and $Y$ -- the~{\it electromagnetic
news functions}. Non-vanishing news functions $d,_u$ and $Y$
correspond to a "rotation" of a radiating source. In vacuum
spacetimes with hypersurface orthogonal Killing vector $\der /\der\f$
we find $d=0$.

The~total  mass of the~system at a given retarded time $u$
is defined by
\BE
m(u)=\pul\int_{0}^{\p} M(u,\th)\sn d\th\ .\label{hmota}
\EE
(Notice that the~definition  of the~total mass given in
\cite{burg} is different;
for example, in the~Schwarzschild case it gives $4\p m_{Schw.}$.)
The~time derivative of the~total mass is equal to
\BEA
m,_u &=&-\pul\int_{0}^{\p}
                (c^*,_u\bar c^*,_u+X^*\bar X^*)\sn d\th
                      \label{klhmota}\\
     &=&-\pul\int_{0}^{\p} (c,_u^2+d,_u^2+X^2+Y^2)\sn d\th
               \leq 0\ . \nn
\EEA
Hence, if any of the~news functions is non-vanishing,
the~waves are radiated out and
the~mass of the~system necessarily decreases.

There exist two quantities which are always conserved:
\BE
\frac{\der}{\der u}\int_{0}^{\p}\pul\ep^*\sn d\th =0\ .
\label{zachnab}
\EE
These  are "the~electric" and "magnetic" charges of the~source.

The~rate of loss of the~electromagnetic energy
radiated out from the~system is given by
\BE
\pul\frac{\der}{\der u}\int_{0}^{\p} X^*\bar X^*\sn d\th\ ,
\label{tokelmgen}
\EE
which also implies the~loss of mass as seen from Eq.~(\ref{klhmota}).

%% file: ajap2.tex
For an illustration we write down the~expression
for the~Lie  derivative component $\L_\e \gnn=0$,
where  the~metric has the~form (\ref{ds}) and
the~vector $\e$ is given by Eqs.~(\ref{Btetrada}), (\ref{Beta}):
\BEA
\L_\e \gnn&=&-\frac{1}{2r^3\co^{\frac{3}{2}}}e^{-3\g-2\b}\nn\\
          & &\ \lvvsz 2r^2\co\edbg
                \lvhz-\sqrt{2ch^2-1}\eg(2A,_u r+B,_u V\edb)\nn\\
              & &\mv\mv\mm
                    -2\edg(2\tilde{g},_u shch+\tilde{f},_u)Ur^2
                          -\co 2r^2\tilde{g},_u W\pvhz\nn\\
              & &\
                   +\tilde{g} 4r^4\edbg
                  \lvhz W\co+2Ushch\edg\pvhz
                    [2\du shch-\co\gu ]\nn\\
              & &\ +\tilde{f} 2r\edb
                   \lvhz 2r^3W\co^2
                  \lvkz 2W\dth shch+\co(\Wth+\gth W)\pvkz
                  -\edbg\co^2(\Vth+2\bth V)\nn\\
              & &\ \  +4r^3\edg\co\lvkz\co^2UW\dth
                  +\co shch(\Uth W+U\Wth+2W\gu)+W\du\pvkz\nn\\
              & &\ \  +2Ur^3 e^{4\g}
                    \lvkz\co^3(\Uth+U\gth) +\co(2\co^2-1)\gu\nn\\
              & &\mv\mv\mv  +2U\dth shch\co^2
                        +2\du shch\pvkz\pvhz\nn\\
              & &\ +A2r^2\eg\co^{\frac{3}{2}}
                     \lvhz 2r^2W\lvkz\co(-rW\gr+W+r\Wr)
                            +2rW\dr shch\pvkz\nn\\
              & &\ \ +2r^2U e^{4\g} \lvkz\co(rU\gr+U+r\Ur)
                     +2rU\dr shch\pvkz+\edbg(-\Vr+\V-2V\br+4r\bu)\nn\\
              & &\ \  +4r^2\edg \lvkz\co rUW\dr
                         +shch(r\Ur W+2UW+rU\Wr)\pvkz\pvhz\nn\\
              & &\ +Be^{2\b+\g}\co^{\frac{3}{2}}\lvhz
                  \edbg\lvkz Vr(\Vr-\frac{1}{r}V+2V\br)-4Vr^2\bu
                      -2Ur^2(2\bth V+\Vth)\pvkz\nn\\
              & &\ \ +2r^4e^{4\g}
                      \lvkz\co\lkz 2U^2r(\Uth+U\gth+\gu)
                       -VU(\frac{1}{r}U+\Ur+U\gr)\pkz
                   +2U^2rshch(2U\dth-\V\dr+2\du)\pvkz\nn\\
              & &\ \  +4r^4\edg\lvkz
                       UW\co(-V\dr+2Ur\dth+2r\du)\nn\\
              & &\mv\mv\mv
                      +shch\lkz -V(\Ur W+U\Wr+\frac{2}{r}UW)
                        +2Ur(\Uth W+U\Wth)\pkz\pvkz\nn\\
              & &\ \ +r^4\lvkz\co 2W
                      \lkz -2r\gu W-\V(W+\Wr r-W\gr r)
                   +2Ur(\Wth-\gth W)\pkz\nn\\
              & &\mv\mv  +4W^2 shch(2Ur\dth-V\dr+2r\du)\pvkz\pvhz
                       \pvvsz=0\nn\ .
\EEA
The~Killing equation  $\L_\e \gnn$ is
the~lengthiest among all Killing equations
$\L_\e \gmn=0$.

%% file: ajapjibi.tex
In paper \cite{bicak} a number of misprints appear,
and there is also a sign error which does not change
the~conclusions of the~paper but makes wrong
Eqs. (53)-(57), including the~pathological solution
given by Eq. (57) for $\e^u$. Below we give the~correct
forms of all equations from \cite{bicak} in which
a misprint appears. First we give the~correct forms
of erroneous Killing equations from Appendix in
\cite{bicak}. These are (notice that functions $f$, $g$
in \cite{bicak} are denoted as $\tilde{f}$, $\tilde{g}$
in the~present paper):
\BEA
\L_\e \gnj&=&\edb\bth (2\tilde{f}r^{-1}\emg+2BU)+(B\edb),_u
           +U\eg(r\tilde{f},_r-\tilde{f}-\tilde{f}r\gr)\\
           &\ &+Ur^2\edg B\Ur
          + \lvkz A-\frac{1}{2r}BV\edb\pvkz,_r
           +\frac{1}{r}B,_r V\edb\ ,\nn\\
\L_\e \gjj&=&2\edb B,_r\ ,\\
\L_\e \gjt&=&\emg\sn(\tilde{g}-\tilde{g}r\gr-r\tilde{g},_r)
                 +\edb B,_\f\ ,\\
\L_\e \gdd&=&-2r\eg \lvhz \tilde{f},_\th+rB\eg\gu+BUr\eg\gth+Br\eg\Uth
             +\eg \ABV(1+r\gr)\pvhz\ .
\EEA
In the~main text in \cite{bicak} the~following equations
contain misprints and are here corrected:
\BEA
\mbox{Eq.}\ (34)&:&\ B,_u+2\An,_u=0\ ,\\
\mbox{Eq.}\ (46)&:&\ \fj,_\th+B,_{\th\th}c+\Aj-c(\AnmB)+BM-B\cc,_\th=0\ ,\\
\mbox{Eq.}\ (47)&:&\ \fj\ctg+B,_\th (c,_\th+c\ctg)+c(\AnmB)
                           +\Aj+BM-B\ctg\cc\ ,\\
\mbox{Eq.}\ (48)&:&\ [B,_\th\ctg+Bc,_u-(\AnmB)]c=0\ .\label{jibi48}
\EEA
%Then using Eqs. (45) and (43)  the Eq. (48) will have a form
%\BE
%[B,_\th\ctg+Bc,_u-(\AnmB)]c=0\ ,
%\EE
%which using (37) imply
%\BE
%[B,_{\th\th}+B,_\th\ctg+B-2\An]c=0\ .
%\EE
As with Eq. (48) in \cite{bicak} (which contains a sign error),
the~correct equation (\ref{jibi48}) implies that either $c=0$
or the~expression in the~square brackets  has to vanish.
The~first possibility gives vanishing news function and leads
to translations. The~second possibility leads to
\BE
\An=\pul(B,_{\th\th}+B,_\th\ctg+B)\ ,
\EE
i.e. to Eq. (\ref{tAn}) of the~present paper. As it is shown
below equation (\ref{tAn}) in the~main text, this case corresponds
to the~translational Killing vector under the~presence
of a straight cosmic string along $z$-axis.

%% file: ajapstru.tex
In this Appendix we show that function $B$ in Eq. (\ref{strB})
corresponds to translations along $z$-axis and $t$-axis
in the~spacetime of an infinite thin cosmic string.
Let us first recall some results from Ref.~\cite{Bstruna}.
The~metric outside a straight non-rotating cosmic string along $z$-axis
can be written in cylindrical coordinates as
\BE
ds^2=dt^2-d\r^2-dz^2-{\rm C}^2\r^2d\f^2\ ,\label{metrikastruna}
\EE
where $\f\in[0,2\p)$ and ${\rm C}\in(0,1]$ is a constant.
(Eq. (\ref{metrikastruna}) represents also
the~asymptotic form of the~metric corresponding
to a spatially bounded system and the~cosmic string along
$z$-axis.)
Introducing spherical flat-space coordinates
$\{R,\ \vth,\ \f\}$ by
$\r=R \sin\vth$, $z=R\cos\vth$, $\f=\f$,
and flat-space retarded time $U=t-R$, we get
\BE
ds^2=dU^2+2dUdR-R^2(d\vth^2+{\rm C}^2\sin^2\vth d\f^2)\ .
\EE
We can now go over to Bondi's coordinates 
$\{u,\ r,\ \th,\ \f\}$ -- in which
the~metric has asymptotically  Bondi's form --
% (5), (6) in \cite{Bstruna} such that
by assuming expansions
\BEA
U   &=&\stackrel{o}{\p}(u,\th)+O(r^{-1})\ ,\nn\\
R   &=&q(u,\th)r+O(r^{0}) \ ,\\
\vth&=&\stackrel{o}{\t}(u,\th)+O(r^{-1})\ .\nn
\EEA
%(see (16) in \cite{Bstruna}). Then the news function reads
%\BE c,_u= \EE
%(see (24) in \cite{Bstruna}). Using (17)-(21) in \cite{Bstruna}
%we can arrive at an expression:
%\BE
%c,_u=\frac{1}{2q^2}(2qq,_{\th\th}-q,_\th^2+q^2-1)\ .
%\EE
%And employing then (22) in \cite{Bstruna}:
Comparing then the~resulting metric with the~general
form of Bondi's metric  one obtains the~expressions
for functions $\stackrel{o}{\p}$, $q$, $\stackrel{o}{\t}$, ...
(see \cite{Bstruna} for details) and the~expression
for the~news function $c,_u$.
Here we need only the~following results:
\BE
q(\th)=\frac{\sn}{{\rm C}\sin\vth}\ ,
\  \stackrel{o}{\t},_\th=\pm\frac{1}{q}\ ,
\EE
and the~news function is
\BE
c,_u=\frac{{\rm C}^2-1}{2\dsn}\ .
\EE
%(25) in [\cite{Bstruna}]
Taking the~$\th$-derivative of $q$, using
$\stackrel{o}{\t},_\th=\pm 1/q$,
and excluding the~original flat-space $\vth$, we obtain equation
\BE
q,_\th^2-2qq,_\th\ctg+q^2\lvkz\dctg-\frac{{\rm C}^2}{\dsn}\pvkz +1=0\ ,
\EE
which implies
\BE
q,_\th=q\ctg\pm\sqrt{\frac{q^2{\rm C}^2}{\dsn}-1}\ .
\EE
The~solution is
\BE
q=\frac{\sn}{2{\rm C}}
       \lvhz\lvkz\chi\frac{\sn}{\cs+1}\pvkz^{\rm C}+
            \lvkz\chi\frac{\sn}{\cs+1}\pvkz^{-\rm C}\pvhz\ ,\
             \chi=\mbox{const}\ .\label{q}
\EE

Consider now translations along $z$-axis and $t$-axis.
In coordinates $\{t,\ \r,\ z,\ \f\}$ they have the~form
$\z_{(z)}^\m=[0,0,a_0,0]$ and $\z_{(t)}^\m=[b_0,0,0,0]$,
with $a_0$ and $b_0$ constant,
so that their general combination is just
$\z_{(z)}^\m+\z_{(t)}^\m$.

The~asymptotic form of this linear combination in
Bondi's coordinates reads
\BEA
\z_{(z)}^u  +\z_{(t)}^u   &=&
      b_0q+\frac{a_0}{{\rm C}}\dsn\lvkz\frac{q}{\sn}\pvkz,_\th+O(r^{-1})\
,\nn\\
\z_{(z)}^r  +\z_{(t)}^r   &=& \frac{b_0}{2q}
            \lvkz {q,_\th}^2-q^2+1\pvkz
           +\frac{a_0}{2{\rm C}q^2}\lvhz
               (-{q,_\th}^2+q^2+1)(q\cs-q,_\th\sn)+2q,_\th\sn\pvhz
         +O(r^{-1}) \ ,\\
\z_{(z)}^\th+\z_{(t)}^\th &=& -\frac{b_0 q,_\th}{r}
              +\frac{a_0}{{\rm C}qr}\lvhz
               q,_\th(q\cs-q,_\th\sn)-\sn\pvhz +O(r^{-2})\ ,\nn \\
\z_{(z)}^\f +\z_{(t)}^\f  &=& 0\ .\nn
\EEA
Substituting for $q$ from Eq. (\ref{q}) we obtain
\BEA
\z_{(z)}^u+\z_{(t)}^u&=&
 \frac{\sn}{2{\rm C}}\lvkz\chi\frac{\sn}{\cs+1}\pvkz^{\rm C} (b_0+a_0)
+\frac{\sn}{2{\rm C}}\lvkz\chi\frac{\sn}{\cs+1}\pvkz^{-\rm C}(b_0-a_0)\nn\\
&=&\sn\lvkz\frac{\sn}{\cs+1}\pvkz^{\rm C}
     \frac{(b_0+a_0)\chi^{\rm C}}{2{\rm C}}
  +\sn\lvkz\frac{\sn}{\cs+1}\pvkz^{-\rm C}
     \frac{(b_0-a_0)\chi^{-\rm C}}{2{\rm C}}\ ,
\EEA
and similar expressions for $\z_{(z)}^r  +\z_{(t)}^r$ and
$\z_{(z)}^\th  +\z_{(t)}^\th$; these  are somewhat
lengthy and are thus not written here explicitly.
Comparing now the~last results (including those for
$\z_{(z)}^r  +\z_{(t)}^r$ and $\z_{(z)}^\th  +\z_{(t)}^\th$)
with a general asymptotic form
of the~Killing vector representing a supertranslation,
i.e. with Eq.~(\ref{etasupertrans})
with $B$ given by Eq. (\ref{strB}), we find that this Killing
vector is just equal to $\z_{(z)}^\m+\z_{(t)}^\m$
if the~constant parameters are related by
\BEA
a&=&\frac{(b_0+a_0)\chi^{\rm C}} {2{\rm C}}\ ,\\
b&=&\frac{(b_0-a_0)\chi^{-\rm C}}{2{\rm C}}\ .
\EEA
Therefore the~supertranslational Killing vector is in fact
the~translational Killing vector in the~asymptotically flat
spacetime with an  infinite cosmic string along $z$-axis;
$a+b=0$, $b_0=0$ corresponds to a translation along $z$-axis,
and $a-b=0$, $a_0=0$~-~along $t$-axis.

For the~special case ${\rm C}=1$ (no string)
the~spacetime is asymptotically flat, and
\BE
q=\cs\lvkz\frac{1}{2\chi}-\frac{\chi}{2}\pvkz
         +\frac{1}{2\chi}+\frac{\chi}{2}
\EE
corresponds to a standard translation where $B$
is given by Eq. (\ref{trB}).
%\BE
%B=\cs \frac{-\chi(b_0+a_0)+\frac{1}{\chi}(b_0-a_0)}{2}
%     +\frac{\chi(b_0+a_0)+\frac{1}{\chi}(b_0-a_0)}{2}\ .
%\EE

%% file: ajapelmg.tex
%\centerline{\Large\bf{Electromagnetic field in the supertranslational case}}
\subsection{Translational case}

The~electromagnetic tensor   reads:
\BEA
\Fnj  &=&-\ep_1 (\th)\frac{1}{r^2}
       + \lvhz uB^{-2}(-\ep_1 BB,_{\th\th}+\ep_1 B,_\th^2
                -\ep_1 BB,_\th\ctg-\ep_1,_\th BB,_\th)
             +e_1,_\th+e_1\ctg\pvhz  \frac{1}{r^3}+O(r^{-4})\ ,\\
         %\label{tFnj}\\
\Fnd  &=&\lvhz\ep_1,_\th+\ep_1 B,_\th B^{-1}\pvhz\frac{1}{r}
             +O(r^{-2})\ ,\\
\Fnt  &=&-\m_1 \sn B,_\th B^{-1} \frac{1}{r}+O(r^{-2})\ ,\\
\Fjd  &=&\lvhz-\ep_1 B,_\th B^{-1}u+e_1\pvhz\frac{1}{r^2}
              +O(r^{-3})\ ,\\%\label{Fjd}
\Fjt  &=&\lvhz\m_1\sn B,_\th B^{-1}u+f_1\sn\pvhz\frac{1}{r^2}
           +O(r^{-3})\ ,\\
\Fdt  &=&-\m_1\sn-
          \lvsz B^{-2}\lvhz (\m_1\sn),_\th BB,_\th
          +\m_1\sn BB,_{\th\th}-\m_1\sn B,_\th^2\pvhz u
                 +(f_1 \sn),_\th\pvsz\frac{1}{r}+O(r^{-2})\ .
\EEA

\subsection{Boost case}
Assuming the~boost-rotation symmetry,
the~electromagnetic tensor  has the~form:
\BEA
\Fnj  &=&-\frac{\E}{u}\ctg\frac{1}{r^2}
        +\lvhz\frac{\ctg}{2}  (\E,_w w+\E)
              +\frac{\ctg}{u^2}(\GG,_w w+\GG)
              %+lu\frac{1+\dcs}{\csn}
              \pvhz\frac{1}{r^3}+O(r^{-4})\ ,\label{bFnj}\\
\Fnd  &=&\frac{\E}{u^2}+
        \lvhz\frac{1}{u\dsn}(\E,_w w-\E)-\frac{\E,_w\sn}{2u^2}
               %-\frac{l\cs}{\tsn}
               +\frac{1}{u^3}(\GG,_w w+2\GG)\pvhz\frac{1}{r}
             +O(r^{-2})\ ,\label{bFnd}\\
\Fnt &=&\frac{\BB}{u}+\lvhz -\pul (\BB,_w w-\BB)
        +\frac{1}{u^2}(\D,_w w+\D)
         %-\frac{p\cs}{\dsn}
        \pvhz\frac{1}{r}+O(r^{-2})\ ,\label{bFnt}\\
\Fjd &=&\lvhz\pul\E
          %-\frac{lu\cs}{\tsn}
        +\frac{\GG}{u^2}\pvhz\frac{1}{r^2}+O(r^{-3})\ ,\label{bFjd}\\
\Fjt &=&\lvhz -\pul\BB u
             %+\frac{pu\cs}{\dsn}
        +\frac{\D}{u}\pvhz\frac{1}{r^2}+O(r^{-3})\ ,\label{bFjt}\\
\Fdt &=&-\BB\ctg%-\frac{p}{\sn}
       +\cs\lvkz\pul\BB,_w-\frac{\D,_w}{u^2}\pvkz
          %+\frac{pu}{\sn}(1+\frac{1}{\dsn})
                   \frac{1}{r}+O(r^{-2})\ .\label{bFdt}
\EEA

Let us yet illustrate the~general asymptotic forms (\ref{bFnd}),
(\ref{bFnt}) for boost-rotation symmetric electromagnetic fields
by specific examples. As follows from Eqs.~(\ref{bFnd}),
(\ref{bFnt}), the~first non-vanishing terms are given by
$\Fnd=\E(w)/u^2$, $\Fnt=\BB(w)/u$, with $w=\sn/u$.
Denote by $e$, $p$, $m$ electric charge, electric dipole and
magnetic dipole moments respectively, and by $\a^{-1}$
the~magnitude of acceleration.
For a uniformly accelerated
electric monopole (producing Born's solution), we find
$\E(w)=e\a^2 w/(1+\a^2 w^2)^{3/2}$, $\BB(w)=0$.
For a uniformly accelerated
electric dipole (see \cite{BiMu} for the~complete field), we get
$\E(w)=p\a w(2-\a^2 w^2)/(1+\a^2 w^2)^{5/2}$, $\BB(w)=0$, whereas for
a magnetic dipole (see \cite{V} for the complete field), one finds
$\E(w)=0$, $\BB(w)=-\a mw^2(2-\a^2 w^2)/(1+\a^2 w^2)^{5/2}$.
We can easily check that $\int(\E\cs /u) d\th=0$ in the~case of electric
dipole and $\int\BB\ctg d\th =0$ in the~magnetic case -
total charges indeed vanish.

%% file: main.bbl
\begin{references}
\bibitem{bicakobecne}
J.~Bi\v c\'ak and B.~Schmidt, Phys.~Rev.~D {\bf 40}, 1827 (1989).
\bibitem{bicak}
J.~Bi\v c\'ak and B.~G.~Schmidt, J.~Math.~Phys. {\bf 25}, 600
(1984).
\bibitem{bicakzeleny}
J.~Bi\v c\'ak, Proc.~Roy.~Soc. A {\bf 302}, 201 (1968).
\bibitem{wini}
J.~Bi\v c\'ak, P.~Reilly, and J.~Winicour, Gen.~Rel.~Grav.
{\bf 20}, 171 (1988).
\bibitem{gomez}
R.~Gom\' ez, P.~Papadopoulos, and J.~Winicour, J.~Math.~Phys. {\bf 35},
4184 (1994).
\bibitem{alcu}
M.~Alcubierre, C.~Gundlach, and F.~Siebel, In Abstr. 15$^{\mbox{th}}$
Int. Conf. on Gen. Rel. Grav., Pune 1997, p. 83.
\bibitem{hawking}
S.~W.~Hawking, G.~T.~Horowitz, and S.~F.~Ross, Phys.~Rev.~D {\bf 51},
4302 (1995).
\bibitem{bicakLes}
J.~Bi\v c\'ak, in {\it Relativistic Gravitation and Gravitational
Radiation},
Proc. of the~Les Houches School of Physics,
eds. \mbox{J.-A.}~Marck and \mbox{J.-P.}~Lasota, Cambridge University Press,
Cambridge, p. 67 (1995).
\bibitem{pleb}
J.~Pleba\' nski and M.~Demia\' nski, Ann. Phys. (N. Y.) {\bf 98},
98 (1976).
\bibitem{bondi}
H.~Bondi, M.~G.~J.~van~der~Burg, and A.~W.~K.~Metzner, Proc. Roy.
Soc. A {\bf 269}, 21 (1962);\\
R.~K.~Sachs, Proc. Roy. Soc. A {\bf 270}, 103 (1962).
\bibitem{burg}
M.~G.~J.~van~der~Burg, Proc.~Roy.~Soc. A~{\bf 310}, 221
(1969) gives the~explicit treatment of the~asymptotic behaviour
of the~Einstein-Maxwell fields in the~Bondi-Sachs coordinates.
\bibitem{ABS1}
A.~Ashtekar, J.~Bi\v c\'ak, and B.~Schmidt, Phys.~Rev.~D
{\bf 55}, 669 (1997).
\bibitem{ABS2}
A.~Ashtekar, J.~Bi\v c\'ak, and B.~Schmidt, Phys.~Rev.~D
{\bf 55}, 687 (1997).
\bibitem{Bstruna}
J.~Bi\v c\'ak and B.~Schmidt, Class.~Quantum~Grav.~{\bf 6},
1547 (1989).
\bibitem{MTW}
C.~W.~Misner, K.~S.~Thorne, and J.~A.~Wheeler, {\it Gravitation},
W.~H.~Feeman, San~Francisco (1973).
\bibitem{AstSchm}
A.~Ashtekar and B.~Schmidt, J.~Math.~Phys. {\bf 21}, 862 (1980).
\bibitem{sachs}
We keep the~original notation of Sachs (see Ref. \cite{bondi}),
which is also used in Ref. \cite{bicak}. Today one rather uses
the~Newman-Penrose notation $\{ l^\a,n^\a,m^\a,\bar m^\a\}$
for the~null tetrad. Notice, however, that usually one chooses
$l^r=1$, where $r$ is an affine parameter along an outgoing null
geodesic $\{ u,\th,\f\} =\mbox{const}$, while in Bondi's formalism
$r$ is the~luminosity distance, and $k^r\not= 1$ in general.
\bibitem{aja}
A.~Pravdov\'a, PhD dissertation, Department of Theoretical
Physics, Charles University, Prague 1998.
\bibitem{polyhom}
P.~T.~Chru\' sciel, M.~A.~H.~MacCallum, and D.~B.~Singleton,
Phil.~Trans.~R.~Soc.~London A {\bf 350}, 113 (1995).
\bibitem{BiMu}
J.~Bi\v c\'ak and  R.~Muschall,
Wissenschaftliche Zeitschrift
der Friedrich-Schiller-Universit\H at Jena {\bf 39}, 15 (1990).
\bibitem{V}
V.~Pravda, PhD dissertation, Department of Theoretical
Physics, Charles University, Prague 1998;\\
see also V.~Pravda and A.~Pravdov\' a,
in the Proc. of
"The Week of Postgraduate Students", Faculty of Mathematics and Physics,
Charles University, Prague, 1998, and gr-qc/9806114, 1998.
\end{references}
